\documentstyle[emulateapj,graphics]{article}

\def \etal {{\em et al.}}
\def \eg {{\em e.g.}}
\def \ie {{\em i.e.}}
\def \cf {{\em c.f.}}
\def \etc {{\em etc.}}

\lefthead{Bunker, Moustakas \& Davis}
\righthead{Resolving the Stellar Populations in a $z=4$ Lensed Galaxy}

\begin{document}

\title{Resolving the Stellar Populations in a $z=4$ Lensed
Galaxy\footnote{Based in large part on observations made at the W.~M.~Keck
Observatory}} \author{Andrew J.\ Bunker\altaffilmark{1,2}, Leonidas A.\
Moustakas\altaffilmark{3} \& Marc Davis} \affil{Astronomy Department,
601 Campbell Hall, University of California, Berkeley CA~94720\\ {\tt
email: bunker, leonidas, marc@astron.Berkeley.EDU}}
\altaffiltext{1}{NICMOS Postdoctoral Research Fellow}
\altaffiltext{2}{Present address: Institute of Astronomy, Cambridge
University, Madingley Road, Cambridge CB3\,0HA, UK} 
\altaffiltext{3}{Present address: Department of Astrophysics, University
of Oxford, 1 Keble Road, Oxford OX1\,3RH, UK}

\begin{abstract}
We present deep near-infrared Keck/NIRC imaging of a
recently-discovered $z=4.04$ galaxy (Frye \& Broadhurst 1998). This is
lensed by the rich foreground cluster Abell~2390 ($z\approx 0.23$)
into highly-magnified arcs 3--5\arcsec\ in length.  Our $H$- and
$K'$-band NIRC imaging allows us to map the Balmer\,$+\,4000$\,\AA\
break amplitude. In combination with high-quality archival HST/WFPC\,2
data, we can spatially resolve stellar populations along the arcs. The
WFPC\,2 images clearly reveal several bright knots, which correspond
to sites of active star formation. However, there are
considerable portions of the arcs are significantly redder, consistent
with being observed $\gtrsim 100$\,Myr after star formation has ceased.
Keck/LRIS long-slit spectroscopy along the arcs reveals that the
Ly-$\alpha$ emission is spatially offset by $\approx 1$\arcsec\ from the
rest-UV continuum regions. We show that this line emission is most probably
powered by star formation in neighboring \ion{H}{2} regions, and that
the $z=4$ system is unlikely to be an AGN.
\end{abstract}


\section{Introduction}
\label{sec:intro}

Although $z>3$ galaxies are being found in increasing numbers (\eg,
Steidel \etal\ 1996a,b; Spinrad \etal\ 1998; Dey \etal\ 1998), their
study has been almost exclusively through rest-ultraviolet (UV) flux
redshifted into the optical.  As such, we merely know about the
near-instantaneous star formation rate -- the UV continuum is dominated
by massive short-lived OB-stars, and hence is sensitive to star
formation on the timescale of only $\sim10$\,Myr. From imaging of local
galaxies, morphologies in the UV can be very different from the
appearance in the optical (\eg, O'Connell 1997).  A comprehensive study
of the underlying stellar populations should also utilize longer rest
wavelengths, where the light from older and less massive (but more
numerous) stars may dominate.

A useful spectral feature in the study of stellar populations is the
4000\,\AA\ break, which arises from metal-line blanketing (predominantly
\ion{Fe}{2}) in late-type stars. Even if the stellar population
is only moderately evolved, there will still be the Balmer break,
dominated by an A-star population. At high redshift, these rest-frame
optical breaks are redshifted into the near-infrared.  Spinrad, Dey \&
Graham (1995) report a detection of an `old' stellar continuum
from infrared imaging of a $z=4.25$ radio galaxy, and studies of more
modest redshift galaxies (\eg, LBDS~53W091 at $z=1.55$) suggest the
presence of stars older than $\sim1$\,Gyr, implying immense formation
redshifts (Dunlop \etal\ 1997; Spinrad \etal\ 1997). 

\begin{figure}[ht]
\plotone{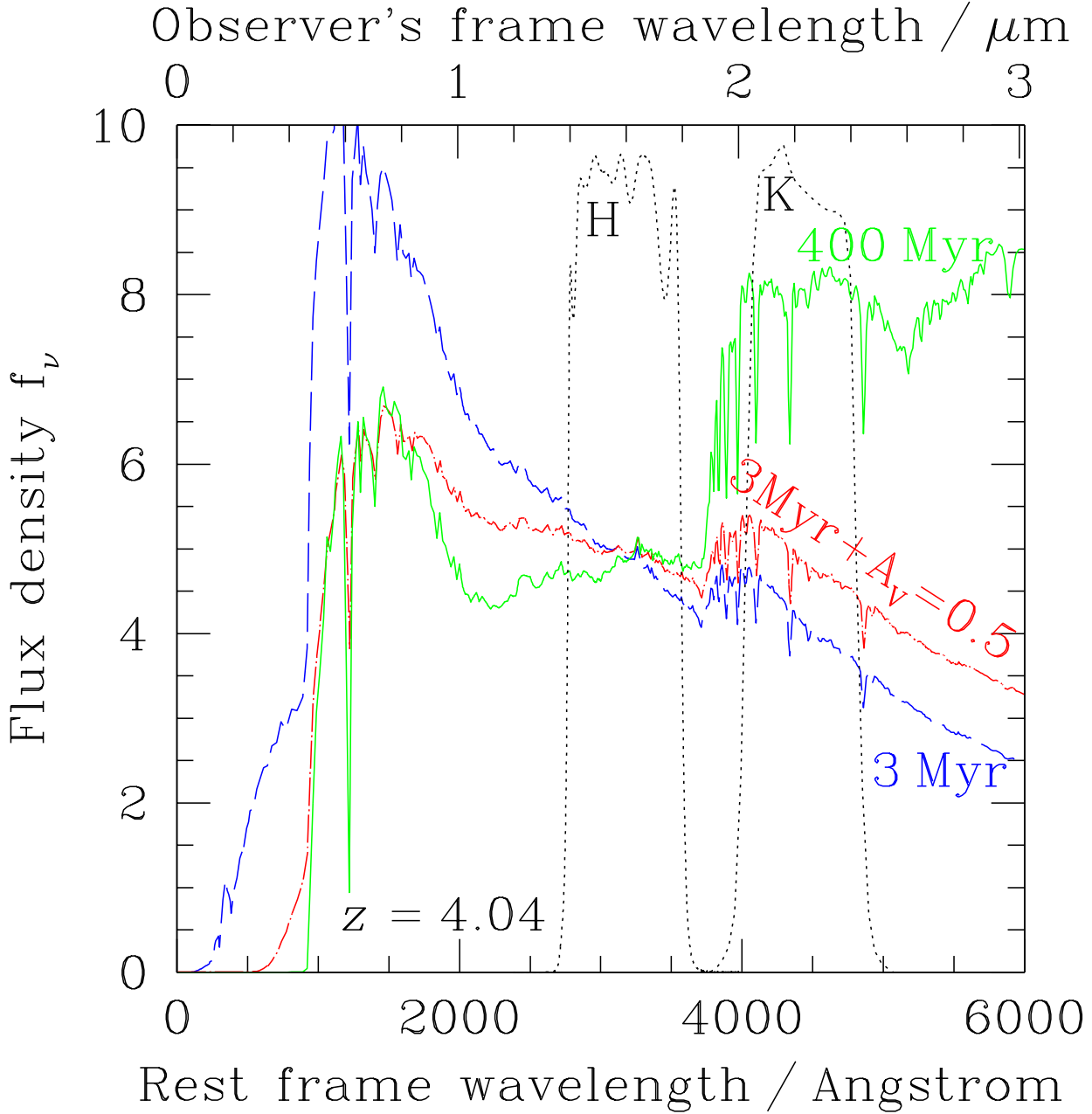} \figcaption[fig1.eps]{An illustration of
the unreddened rest-frame optical spectra of two galaxies, one observed
only 3\,Myr after the end of an instantaneous burst of star formation
(long-dash curve) and the other seen after 400\,Myr have elapsed (solid
line). We also show the 3\,Myr model with dust extinction of
$A_{V}=0.5^{m}$ (using the Calzetti 1997a,b empirical reddening law),
typical of high-$z$ star-forming galaxies (\eg, Pettini \etal\ 1998;
Steidel \etal\ 1999). Note the strong Balmer\,+\,4000\,\AA\ break due to
the older stars. Also plotted (dotted lines) are the $H$ and $K$ filters
in the rest-frame of a $z=4.04$ galaxy, straddling the break. The SEDs
come from the Bruzual \& Charlot models, as updated in Charlot \etal\
(1996).
\label{fig1}}
\end{figure}

\placefigure{fig1}
\placefigure{fig2}

At $z\approx4$, the $H$ ($\lambda_{\rm cent}\approx1.65\,\mu$m) and $K$
($\lambda_{\rm cent}\approx 2.2\,\mu$m) near-infrared (IR) pass-bands
straddle the rest-frame 4000\,\AA\,$+$\,Balmer break
(Figs.~\ref{fig1}\,\&\,\ref{fig2}).  For a spatially-resolved source, it
is possible to map the underlying stellar populations through these
colors, provided this break is not washed-out by a current star-burst.
Here we present an infrared study of stellar populations in a
recently-discovered lensed galaxy at $z=4.04$ (Frye \& Broadhurst 1998,
hereafter FB98).  This system of arcs (Fig.~\ref{fig3}) is among the
highest-redshift gravitationally-magnified galaxies known (see also
Trager \etal\ 1997; Franx \etal\ 1997). These lensed sources are highly
amplified but are not pre-selected for some AGN characteristic or
extreme luminosity, nor are they preferentially color-selected for a
young stellar population.  We will show in \S\,\ref{subsec:noAGN} \&
\S\,\ref{subsec:resonant} that these arcs are likely to be lensed images
of a normal galaxy, affording us the opportunity to study stellar
populations in the early Universe. Hence, we can address whether
galaxies at the highest known redshifts are genuine `proto-galaxies',
observed in the throes of a massive initial starburst, or whether the
star formation history is more extended and we are witnessing `mature'
systems.

\placefigure{fig3}

The structure of this paper is as follows: in \S\,\ref{sec:optical} we
describe the optical imaging with the Hubble Space Telescope (HST), and
\S\,\ref{subsec:lris} presents spatially-resolved optical spectroscopy
obtained with the Low Resolution Imaging Spectrograph (LRIS) on the 10-m
Keck~II telescope. Our near-infrared imaging of these arcs with the Near
Infrared Camera (NIRC) on Keck~I is detailed in \S\,\ref{sec:nirc}. The
broad-band optical/near-infrared colors are used to constrain the dust
extinction and stellar populations in \S\,\ref{sec:disc}, in which we
also address the lensing magnification and the intrinsic luminosity of
this $z\approx 4$ system. Our conclusions are summarized in
\S\,\ref{sec:conc}, and in Appendix~\ref{sec:straight_arc} we present
our near-IR/optical photometry of the nearby ``straight arc'' (Pell\'{o}
\etal\ 1991). Throughout, we assume a cosmology with a vanishing
cosmological constant ($\Lambda_{0}=0$), $H_{0}=50\,h_{50}\,{\rm
km\,s}^{-1}\,{\rm Mpc}^{-1}$ and $q_{0}=0.5$ ($q_{0}=0.1$) unless
otherwise stated.  At $z=4.04$, the Universe is only
$1.2\,h_{50}^{-1}$\,Gyr ($2.1\,h_{50}^{-1}$\,Gyr) old, a look-back time
of 91\% (87\%) the age of the Universe. The proper motion distance is
$d_{M}=6.65\,h_{50}^{-1}$\,Gpc ($d_{M}=11.2\,h_{50}^{-1}$\,Gpc), and in
the absence of magnification, 1\arcsec\ corresponds to a projected
physical distance of $6.4\,h_{50}^{-1}$\,kpc ($10.7\,h_{50}^{-1}$\,kpc).

\section{Lensed Arcs at High Redshift Behind Abell 2390}
\label{sec:optical}

Deep ground-based imaging (\eg, Pell\'{o} \etal\ 1991; Squires \etal\
1996) and more recent HST studies (\eg, B\'{e}zecourt \& Soucail 1997)
of Abell 2390 ($z\approx0.23$) have revealed many giant arcs arising
from gravitational lensing of background galaxies by this massive
cluster, which has a velocity dispersion of $1093\,{\rm km\,s}^{-1}$
(Carlberg \etal\ 1996).  In particular, an $I=18.2^{m}$
elliptical\footnote{Unless otherwise stated, we adopt the natural
Vega-based magnitude system, where the conversion to flux density for a
source of zeroth magnitude is given for the F555W \& F814W HST/WFPC\,2
bands by $f_{\nu}(V_{555}=0)=3664$\,Jy \& $f_{\nu}(I_{814}=0)=2466$\,Jy,
and for the near-infrared filters by $f_{\nu}(J=0)=1578$\,Jy,
$f_{\nu}(H=0)=1020$\,Jy \& $f_{\nu}(K'=0)=688$\,Jy. For reference, the
traditional $K$-band has a zeropoint of $f_{\nu}(K=0)=646$\,Jy on this
system.} close to the critical curve bisects an extended pair of thin,
near-straight arcs as shown in Fig.~\ref{fig3}.  The lensing cluster
elliptical is at $z=0.22437$ (galaxy \#101190 in Yee \etal\ 1996), and
its coordinates are given in Table~1. The arcs are
`stretched' approximately along the major axis of the elliptical, and
the geometry relative to the critical curve suggests a `cusp' arc
(Pell\'{o} \etal\ 1999). The brighter northern arc ($I_{814}=22.6^{m}$)
is $\approx 5$\arcsec\ long, with the southern component
($I_{814}=23.1^{m}$) $\approx 3$\arcsec\ in length.


An arclet redshift survey by B\'{e}zecourt \& Soucail (1997) detected a
single emission line at 6137\,\AA\ from the northern component (their
arclet\,\#20). Subsequent Keck/LRIS spectroscopy by FB98 determined the
redshift of this arc to be $z=4.04$ on the basis of several rest-frame
UV interstellar absorption features (\eg, \ion{C}{2}\,1335\,\AA,
\ion{Si}{4}\,$\lambda$\,1394,\,1403\,\AA\ -- see also
\S\,\ref{subsec:contspec}), consistent with the emission line of
B\'{e}zecourt \& Soucail being Ly-$\alpha$. In fact, Ly-$\alpha$ was
seen in absorption rather than emission in the northern arc by FB98 in
their discovery spectrum; the spectroscopic slit ran perpendicular to
the axis of the arcs, intercepting only $\sim 20$\% of the northern
one. As we show in \S\,\ref{subsec:offsetline}, the Ly-$\alpha$ emission is
concentrated over a small portion of the arcs.  The southern arc and one
counter-arc have also been confirmed to be at this redshift, with
Ly-$\alpha$ in emission (FB98).  Our own spatially-resolved Keck/LRIS
spectroscopy (\S\,\ref{subsec:lris}) reproduces the line emission from
both the northern and southern arcs. The similarity in colors
($B-V>3^{m}$, objects 4\,\&\,5 in Pell\'{o} \etal\ 1991) and their
positions relative to the caustic suggest that these arcs are lensed
images of the same source (see \S\,\ref{subsec:lensgeom}).

To undertake a comprehensive spatially-resolved study of this system of
arcs, we have used archival HST imaging in the optical
(\S\,\ref{subsec:wfpc2}) and long-slit spectroscopy with Keck/LRIS
(\S\,\ref{subsec:lris}) to supplement our near-infrared Keck/NIRC
images, which will be described in \S\,\ref{sec:nirc}.  Throughout, we
have corrected for foreground Galactic extinction using the recent
COBE/DIRBE \& IRAS/ISSA dust maps of Schlegel, Finkbeiner \& Davis
(1998) which have an optical reddening of $E(B-V)=0.11^{m}$, equivalent
to extinctions of $A_{555}=0.33^{m}$, $A_{814}=0.20^{m}$,
$A_{H}=0.06^{m}$ \& $A_{K'}=0.04^{m}$.

\subsection{WFPC2 Observations}
\label{subsec:wfpc2}

We have re-reduced archival HST imaging of the Abell~2390 cluster from
the program of Fort \etal\ (GO-5352) taken with the Wide Field/Planetary
Camera \#2 (WFPC\,2; Trauger \etal\ 1994).  The imaging was undertaken
on 1994 December 10 UT in the F555W and F814W filters, approximating the
ground-based $V$ and $I$ bandpasses.  The data comprised four orbits in
F555W and five in F814W, with each orbit consisting of a single
integration of $2100$\,s, taken at a telescope PA of $+245^{\circ}$.
The telescope was offset by an integer number of pixels between orbits
-- which unfortunately meant that `drizzling' (Hook \& Fruchter 1997) or
other methods for recovering some of the WF undersampling were
ineffective for this dataset.  The data-quality files were used to
create masks of unreliable pixels. When the frames were registered,
these pixels were excluded from the average. The offsets were determined
from the centroid of the star 7\arcsec\ from the lensing elliptical
(Table~1).  Cosmic rays were rejected using the {\tt crreject} algorithm
in {\tt IRAF.IMCOMBINE}; a clipping at the $3\,\sigma$ level was found
to be satisfactory.

For the image in each WFPC\,2 filter we generated isophotal models of
the elliptical using the {\tt STSDAS.ISOPHOTE} package, returning
parameters PA\,$=+35^{\circ}$, $\frac{b}{a}=0.7$ and $r_{e}=1\farcs2$ --
equivalent to $R_{e}=5.5\,h^{-1}_{50}$\,kpc at $z=0.224$.  We then
subtracted this model from each image (Fig.~\ref{fig6}). As noted by
FB98, two smaller arclets close to the elliptical center are revealed,
located on the radial line from the elliptical to the cluster
center. One of these smaller arcs has been spectroscopically confirmed
to be at the redshift of the two main arcs (Frye \etal\ 1997).

For the combined F555W image (8400\,s), the $1\,\sigma$ limiting
magnitude reached in 1\,\sq\arcsec\ was
$V_{555}=28.4$\,mag\,arcsec$^{-2}$, and for the composite F814W frame
($10500$\,s), it was $I_{814}=27.6$\,mag\,arcsec$^{-2}$.  This is
consistent with the predicted sensitivity based on the Poissonian
counting statistics and the WF readout noise ($5\,e^{-}$). The arcs
appear in the WF3 chip, and are clearly resolved along their
length. Careful comparison of transverse cross-cuts of these arcs with
the $V\approx 22^{m}$ star 7\arcsec\ from the elliptical reveals that
the arcs are marginally resolved in their width (Fig.~\ref{fig4}).  The
transverse extent of the arcs is $\approx0\farcs3$ (FWHM), and the WF
point spread function (PSF) has FWHM\,$\approx 0\farcs15$, so the width
(in the absence of any tangential shear) would be $1.8\,h^{-1}$\,kpc
($3.0\,h^{-1}$\,kpc) after deconvolution with the PSF.  For this WFPC\,2
imaging, the $5\sigma$ limiting magnitudes are $V_{555}=26.5^{m}$ and
$I_{814}=25.7^{m}$ using a circular aperture of diameter $0\farcs 5$ to
maximize the signal-to-noise ratio for the characteristic width of the
arcs.

\placefigure{fig4}

Given that the arcs are marginally resolved in the transverse direction
with HST, a na\"{\i}ve model of a spherical source of diameter
$\sim$\,0\farcs3, linearly stretched into the observed arcs 3--5\arcsec\
in length implies an average magnification of $\mu\sim$\,10--17, where
$\mu$ is the integrated flux amplification (in this instance equal to
the linear stretch); to first order, the radial (de)magnification with
respect to the cluster center will be small compared to the tangential
stretching, although the shear stretching produced by the elliptical
galaxy is a significant second-order effect. Although the magnification
may not be constant along the length of the arcs (as we assume in this
simple model), we expect the magnification to be smoothly varying as the
images are some way from the caustic. This estimate of the magnification
is comparable to the value of $\sim 20$ obtained from the lens model of
FB98, who also show that the cluster potential is softer than
isothermal\footnote{In a recent paper, Pell\'{o} \etal\ (1999) use a
different lens model, yielding lower magnifications equivalent to
$8.5\pm2.5$ and $6\pm2$ for the northern and southern arcs (H3a \& H3b
in their notation). This lower magnification is compatible with the
observed morphology if the source is significantly elongated and its
major axis is serendipitously near-coincident with the direction of
linear stretch. In addition to these `cusp' arcs, Pell\'{o} \etal\ also
report the discovery of a `fold' arc (H5) which is located 30\arcsec\
away on the critical curve, but lies outside the region surveyed in our
infrared imaging.}. However, the existence of sub-structure (the
prominent knots N1--N4 and S1--S3 in the northern and southern arcs,
Fig.~\ref{fig3}) indicates that the source object has greater
complexity.  As both main arcs are significantly extended in the
tangential direction, spatially-resolved spectroscopy along the arcs is
possible even with ground-based seeing, and in the next section we
describe our follow-up spectroscopy using Keck/LRIS with a long-slit
aligned along the axis of the northern and southern arcs.

\section{Keck/LRIS Spatially-Resolved Spectroscopy}
\label{subsec:lris}

\subsection{LRIS Observations and Data Reduction}
\label{subsec:lrisobs}

We obtained the long-slit spectroscopy on the night of 1997 July 01 UT
using the Low Resolution Imaging Spectrograph (LRIS; Oke \etal\ 1995) at
the f/15 Cassegrain focus of the 10-m Keck~II Telescope.  The LRIS
detector is a Tek 2048$^2$ CCD with 24\,\micron\ pixels, although the
CCD was windowed-down spatially to just the 900 pixels illuminated by
the 3\arcmin\ by 1\arcsec\ long-slit. The angular scale is 0\farcs
212\,pixel$^{-1}$, and we read out the CCD in two-amplifier mode. The
observations were obtained using the 600~line~mm$^{-1}$ grating in first
order blazed at 7500\,\AA, producing a dispersion of
1.256\,\AA\,pixel$^{-1}$. The reference arc lamps and sky-lines have
FWHM~$\approx$~3.5--5\,pixels (the spectral focus being best at the
central wavelength), so for objects that fill the 1\arcsec-wide slit,
the velocity width of a spectrally unresolved line is FWHM~$\approx
220-325$\,km\,s$^{-1}$. The grating was tilted to sample the wavelength
range $5810-8630$\,\AA, corresponding to $\lambda_{\rm
rest}=1150-1710$\,\AA\ at $z=4.04$. The long-slit observations were
taken with a PA of $+23^{\circ}$ east of north, to cover as much of both
arcs as possible (Fig.~\ref{fig6}). An offset of $\Delta({\rm
RA})=-4\farcs92$, $\Delta({\rm dec})=+1\farcs90$ from a nearby star
(Fig.~\ref{fig3}, with coordinates in Table~1) was performed to center
the arcs in the slit. A total of 3600\,s of on-source integration was
obtained, and this was broken into two individual exposures each of
duration 1800\,s to enable more effective cosmic ray rejection.  The
telescope was dithered by 21\farcs6 along the slit between integrations
to aid in the removal of fringing in the red and the elimination of bad
pixels.  The observations spanned an airmass range of 1.00--1.02. The
seeing had a FWHM~=~0\farcs 6--1\farcs 1 over the course of the night.
Spectrophotometric standard stars HZ~44 \& Feige~110 (Massey \etal\
1988; Massey \& Gronwall 1990) were observed at similar airmass to
determine the sensitivity function, although the night was rendered
intermittently non-photometric by thin cirrus.

Each frame first had the bias subtracted (determined from the overscan
region appropriate to that amplifier), and was converted to electrons
through multiplication by the gain (1.97\,$e^{-}\,{\rm count}^{-1}$ for
the left amplifier, 2.1\,$e^{-}\,{\rm count}^{-1}$ for the right). A
high signal-to-noise dark current frame (formed by averaging many dark
current exposures) was then subtracted.  Contemporaneous flat fields
were obtained with a halogen lamp immediately after the science
exposures, and these internal flats were normalized through division by
the extracted lamp spectrum. Wavelength calibration was obtained from
Ne$+$Ar reference arc lamps, and a fourth-order polynomial fit to the
centroids of 40 arc lines created a wavelength solution with {\em rms}
residuals of 0.07\,\AA. A small ($\approx 1$\,\AA) wavelength zeropoint
correction was applied to the dispersion scale to compensate for drift
in the grating angle between the arc spectra and the data frames; this
was determined by measurement of the centroids of prominent sky
lines. Sky subtraction was performed by fitting a fifth-order polynomial
to each detector column (parallel to the slit), excluding from the fit
those regions occupied by sources.

\subsection{Spectral Continuum Features and Redshift of the Lensed Arcs}
\label{subsec:contspec}

The two-dimensional spectrum shows continuum, spatially distinct from
the light profile of the lensing elliptical. This extended continuum
coincides with the areas occupied by the arcs in the WFPC\,2 imaging
(Fig.~\ref{fig3}).  The continua from the arcs have several absorption
features, and a spectral slope fairly flat in $f_{\nu}$ longward of
$\lambda_{\rm rest}1216$\,\AA\ at $z\approx 4.04$ -- Fig.~\ref{fig5}
shows the one-dimensional spectral extractions. Both the northern and
southern arcs exhibit the same absorption features (see
Table~2), which are consistent with
\ion{O}{1}\,$\lambda$\,1302.2\,\AA\,/\,\ion{Si}{2}\,$\lambda$\,1304.4\,\AA,
\ion{Si}{4}\,$\lambda\lambda$\,1393.8,1402.8\,\AA,
\ion{Si}{2}\,$\lambda$\,1526.7\,\AA\ and
\ion{Fe}{2}\,$\lambda$\,1611.2\,\AA\ at $z=4.039\pm0.004$. Other
rest-UV absorption lines commonly observed in starbursts (such as
\ion{C}{4}\,$\lambda\lambda$\,1548.2,1550.8\,\AA) lie close to prominent
sky lines in the OH forest.  We note that most of these absorption line
wavelengths do not match any common spectral feature at the redshift of
the lensing galaxy ($z=0.22437$), and so the continuum detection from
the region of the arcs cannot be extended light from the foreground
elliptical.  In the continuum of the $z=4.04$ arcs, there are also two
prominent absorption features at 6450\,\AA\ (as noted in FB98) and at
6210\,\AA , which are attributable to foreground absorbers. The
absorption profiles appear to be more consistent with the
\ion{C}{4}\,$\lambda\lambda$\,1548.2,1550.8\,\AA\ absorption doublet at
$z=3.225$ \& $z=3.010$ than with lower-$z$
\ion{Mg}{2}\,$\lambda\lambda$\,2795.5,2802.7\,\AA\ systems.


\placefigure{fig5}

\subsection{Spatially-Offset Line Emission Associated with the $z=4$ Arcs}
\label{subsec:offsetline}

Fig.~\ref{fig6} shows the area of the WFPC\,2 image intercepted by the
long-slit, displayed alongside the two-dimensional spectrogram (trimmed
to the spectral region around Ly-$\alpha$ at $z=4.04$).  Our optical
spectroscopy reveals two sites of prominent line emission, detected at
high signal-to-noise (${\rm SNR}\approx 70$).  However, as is readily
apparent from Fig.~\ref{fig6}, only one of the four bright knots of the
northern arc (labeled N4 in Fig.~\ref{fig3}) has a prominent nearby
emission line, and similarly only S1 in the southern component is close
to a region of line emission. However, neither of these knots detected
in continuum in the WFPC\,2 images is coincident with the centroids of
the emission lines seen in the long-slit spectroscopy, which lie to the
north of the continuum in both cases (Table~1). The centroid of the
northern line emission is $8\farcs7\pm0\farcs2$ from the center of the
elliptical (resolved along the spectroscopic slit axis of PA
$23^{\circ}$), but the centroid of the nearest continuum knot, N4, is
$7\farcs6\pm0\farcs1$ from the elliptical in the WFPC\,2 images, with
the northern extreme of the continuum extending $<0\farcs5$ beyond.
Hence the spatial offset between the continuum and the center of the
northern line emission is larger than the size of the PSFs and much
greater than the uncertainty in centroiding.  With its smaller
magnification, this spatial incoherence is less pronounced in the
southern arc: the Ly-$\alpha$ emission is $3.7\pm0\farcs2$ from the
center of the elliptical, but the centroid of the nearest continuum knot
(S1) is further from the elliptical ($4\farcs3\pm0\farcs1$).

\placefigure{fig6}

The spatially-offset line emission near S1 has a central wavelength of
$(6134\pm 1)$\,\AA , with the error dominated by the $\approx 1$\,pixel
(1.256\,\AA) uncertainty in centering the emission line in the slit.
The wavelength is consistent with Ly-$\alpha$\,$\lambda$\,1215.7\,\AA\
at $z=4.0456\pm0.001$, within 400\,km\,s$^{-1}$ of the absorption-line
redshift for the continuum. We first consider and reject the unlikely
scenario where the line emission is not Ly-$\alpha$ arising from the
$z=4.04$ object responsible for the continuum of the arcs, but instead
is a physically distinct source. Not only would this have to be a chance
projected alignment (to within $0\farcs1\times({\mu}/{10})^{-1}$ in the
source plane), but also by coincidence the physically-distinct source
would have to have a redshift such that its solitary emission line had a
wavelength mimicking Ly-$\alpha$ at the redshift of the arcs to within
$\frac{\Delta\lambda}{\lambda}\approx 0.001$. As any line at a shorter
rest-frame wavelength than 1216\,\AA\ would be effectively extinguished
by the Ly-$\alpha$ forest absorption, an object at $z<4$ would be
required. This source would have to be extremely unusual in being
capable of producing an isolated emission line with huge equivalent
width: no continuum is detected at the location of the line
emission. The 3\,$\sigma$ lower limit on the equivalent width is $W_{\rm
obs}>500$\,\AA\ in the observer's frame, derived from the noise in the
F555W image within a $1\sq\arcsec$ aperture centered at the location of
the line emission. This line emission is extremely unlikely to be
H$\beta$\,$\lambda$\,4861.3\,\AA\ at $z=0.262$ as we do not detect
H$\alpha$\,$\lambda$\,6562.8\,\AA\, which would be $>3$ times as strong
(assuming case~B recombination). Similarly,
[\ion{O}{3}]\,$\lambda$\,5006.8\,\AA\ at $z=0.225$ is ruled out:
coincidentally, this redshift is close to that of the lensing
elliptical, but the absence of the second line in the [\ion{O}{3}]
doublet makes this interpretation untenable --
[\ion{O}{3}]\,$\lambda$\,4958.9\,\AA\ is one third as strong (Osterbrock
1989), well above our detection threshold. The only other remotely
plausible interpretation for the line emission is
[\ion{O}{2}]\,$\lambda\lambda$\,3726.1,3728.9\,\AA\ at $z=0.646$;
however, our spectrogram extends out to H$\beta$ \& [\ion{O}{3}] for
this redshift, which are not observed. Even if the absence of these
lines was on account of anomalous line ratios, we would still expect to
marginally resolve the [\ion{O}{2}] doublet (although any velocity
smoothing greater than about 250\,km\,s$^{-1}$ will wash out the
splitting). Indeed, our spectroscopic long-slit did serendipitously
intercept a line-emission galaxy at $z=1.129$ (indicated in
Fig.~\ref{fig6}) 3\arcsec\ below the southern arc (Table~1), which
clearly exhibits the characteristic doublet profile of [\ion{O}{2}],
with the redshift confirmed by [\ion{Ne}{3}]\,$\lambda$\,3868.7\,\AA\
(Fig.~\ref{fig7}).  We do not see this doublet profile in the line
emission adjacent to the arcs. Hence we conclude that Ly-$\alpha$
emission, associated with the continuum from the $z=4.04$ arcs, is by
far the most probable identification of the line. This interpretation
also has the advantage that there exists a nearby source of UV photons
(in the adjacent star-forming knots, N4 \& S1) with which to generate
the Ly-$\alpha$ recombination line; in \S\,\ref{subsec:resonant} we
describe a model for the transport of these photons from the
star-forming regions to the areas observed in line emission.

\placefigure{fig7}

\subsection{Spectral Profile of the Ly-$\alpha$ Emission}
\label{subsec:lineprof}

The Ly-$\alpha$ line is clearly resolved, with a measured spectral width
of 7.8\,pixels FWHM for the S1 knot, which was centered on the
slit. This is equivalent to a velocity width of $\Delta v_{\rm
FWHM}=370\,{\rm km\,s}^{-1}$ after deconvolution with the instrumental
width -- really a lower limit as the source did not quite fill the
1\arcsec-wide slit in the sub-arcsec seeing. The profile is asymmetric,
with a pronounced red wing but a sharper decline in flux density on the
blue side (Fig.~\ref{fig8}). This appears to be a common feature in
high-$z$ starbursts with Ly-$\alpha$ in emission (\eg, Dey \etal\ 1998;
Lowenthal \etal\ 1997), and is most likely due to an outflow of neutral
hydrogen, where we only see the back-scattered Ly-$\alpha$ from the far
side of the expanding nebula -- only the photons on the red side of the
resonant Ly-$\alpha$ emission-line profile can escape, with the blue
wing being absorbed by neutral gas (mainly within the galaxy). This also
accounts for the $400\,{\rm km\,s}^{-1}$ difference in redshift between
that determined from the Ly-$\alpha$ emission (the profile of which has
been skewed to the red) and the redshift measurement from the absorption
features (which may be more representative of the systemic redshift,
although large-scale outflows in the ISM may result in a relative
blueshift of these absorption lines). A spectral extraction encompassing
the continuum and line emission regions shows that there is Ly-$\alpha$
absorption blueward of the emission line (Fig.~\ref{fig8}). The
P\,Cygni-like profile is consistent with this outflow model. Recent
near-IR spectroscopy by Pettini \etal\ (1998) of the rest-frame optical
forbidden lines and Balmer lines in $z\approx 3$ galaxies also shows a
$> 300\,{\rm km\,s}^{-1}$ redshifting of the Ly-$\alpha$ relative to the
nebular emission lines.

\placefigure{fig8}

The velocity dispersion of $\sigma = 160\,{\rm km\,s}^{-1}$ inferred
from the line-width of Ly-$\alpha$ is likely to be a lower limit, as the
line profile is truncated. However, the velocity width of Ly-$\alpha$ is
unlikely to be representative of the velocity dispersion of the galaxy,
as resonant scattering will broaden the spectral profile of the escaping
photons (\eg, Binette \etal\ 1993).  Spatially, the two regions of
Ly-$\alpha$ emission are resolved along the slit (the major axis of the
arcs) and extend over 1--2\arcsec\ FWHM.  The flux in the line from S1
is $f=37\times 10^{-18}\,{\rm ergs\,s^{-1}\,cm^{-2}}$ (N4 falls on the
edge of the slit and is presumably subject to significant slit
losses). This is equivalent to a line luminosity of $0.68\times
10^{42}\,(\mu/10)^{-1}\,h^{-2}_{50}$\,ergs\,s$^{-1}$ ($1.9\times
10^{42}\,(\mu/10)^{-1}\,h^{-2}_{50}$\,ergs\,s$^{-1}$), which in the
absence of extinction of this line corresponds to a star formation
rate\footnote{We convert the line flux to a star formation rate using
the Bruzual \& Charlot (1993) spectral evolutionary models, as updated
in GISSEL 1996 (Charlot, Worthey \& Bressan, 1996) with a Salpeter
(1955) IMF of mass range $0.1\,M_{\odot}<M_{*}<125\,M_{\odot}$.  From
this, a constant star formation rate of $1\,M_{\odot}\,{\rm yr}^{-1}$
will produce $\Psi_{\rm LyC}=1.44\times 10^{53}\,{\rm s}^{-1}$ of Lyman
continuum photons ($\lambda_{\rm rest}<912$\,\AA).  Assuming case~B
recombination, 69\% of these Lyman continuum photons will be reprocessed
as Ly-$\alpha$ (Spitzer 1978, table~1). The number of Lyman continuum
photons per unit star formation rate is a factor of $\approx 2$ higher
(so the inferred SFR a factor of $\approx 2$ lower) than that obtained
from the Kennicutt (1983) IMF.}  of
$0.42\,(\mu/10)^{-1}\,h^{-2}_{50}\,M_{\odot}\,{\rm yr}^{-1}$
($1.1\,(\mu/10)^{-1}\,h^{-2}_{50}\,M_{\odot}\,{\rm yr}^{-1}$).  This
recombination line emission could be generated by the unobscured UV flux
of about $10^4\,(\mu/10)^{-1}\,h^{-2}_{50}$ ($3\times
10^4\,(\mu/10)^{-1}\,h^{-2}_{50}$)~O7-stars.

\label{subsec:noAGN}

Careful inspection of the spectrum did not reveal any emission lines at
other wavelengths, either in the region of the continuum or in the areas
occupied by the offset Ly-$\alpha$ emission. Table~3 gives the upper
limits on the flux in high-ionization rest-UV lines falling within our
spectral coverage, comparing with the typical line ratios from composite
AGN spectra of QSOs (Osterbrock 1989), radio galaxies (Stern \etal\
1999) and Seyfert~IIs (Ferland \& Osterbrock 1986). The non-detections
of \ion{N}{5}\,1240\,\AA, \ion{C}{4}\,1549\,\AA\ \&
\ion{He}{2}\,1640\,\AA\ strongly favor the interpretation that the
Ly-$\alpha$ arises from the Lyman continuum flux produced by OB stars,
rather than the harder UV spectrum of an AGN.


\subsection{Lens Geometry}
\label{subsec:lensgeom}

It is likely that the northern and southern arcs are images of the same
source, and the lens geometry suggests these have even parity (FB98).
The locations where there is distinct Ly-$\alpha$ emission (at the
northern-most tip of each arc) corroborate this. A further test comes
from the Ly-$\alpha$ line centroids measured in our high-dispersion
optical spectroscopy; if the interpretation of these arcs as lensed
images of the same source is correct then we would expect the central
wavelength of Ly-$\alpha$ to be identical; alternatively, two associated
galaxies at the same redshift would presumably exhibit a velocity
difference of $\Delta v > 100\,{\rm km\,s}^{-1}$ if gravitationally
bound. In fact, our slit PA was such that although S1 was centered on
the slit, N4 was off-center by 2\,pixels (Fig.~\ref{fig6}) -- the arcs
are not exactly co-linear, so a compromise slit position was used to
encompass as much of both arcs as possible.  Allowing for this spatial
offset of 0\farcs4, the line centroids of the Ly-$\alpha$ from N4 \& S1
are identical to within our measurement error ($\Delta v<50\,{\rm
km\,s}^{-1}$) as illustrated in Fig.~\ref{fig9}. The relative photometry
of both arcs from the deepest WFPC\,2 F814W image (Table~4) 
suggests that the ratio of the magnification of the northern arc to the
southern is 1.6:1 ($\Delta m = 0.5^{m}$). This is consistent with the
relative lengths of the arcs (5\arcsec\,:\,3\arcsec). Unfortunately, the
proximity of the northern-most portion of the northern arc to the edge
of the spectroscopic long-slit (Fig.~\ref{fig6}) implies that the
northern Ly-$\alpha$ emission-line region is probably subject to large
slit losses, so we cannot use the measured line flux ratio between north
and south as a further independent check on the relative magnification.

\placefigure{fig9}

\section{A Near-Infrared Study}
\label{sec:nirc}

\subsection{Observations with Keck/NIRC}
\label{subsec:obs}

For our near-infrared imaging, we used the Near-Infrared Camera (NIRC;
Matthews \& Soifer 1994) at the f/25 forward Cassegrain port of the 10-m
W.M.~Keck~I Telescope.  NIRC is a 256$^2$ InSb array with an angular
scale of $0\farcs15$\,pixel$^{-1}$.  The observations were undertaken on
1997 August 25 UT in photometric conditions.  Although the seeing in the
near-infrared was $\lesssim 0\farcs5$ throughout the night, an intermittent
drift in the telescope focus rendered the PSF of some frames
significantly worse.

A concern with using broad-band magnitudes to infer the continuum levels
is possible contamination by emission lines (\eg, Eales \& Rawlings
1993).  We used the $K'$ filter ($\lambda_{\rm cent}=2.1\,\mu$m), which
at the redshift of the target galaxy lies red-ward of the 4000\,\AA\
break, but avoids any possible contamination from the
H$\beta$\,4861.3\,\AA\ recombination line at $\sim 2.445\,\mu$m, unlike
the usual $K$-band which has a longer-wavelength cut-off. The $K'$-band
is also a good approximation to the rest-frame $B$-band.

We observed Abell~2390 in both $H$ ($2280$\,s) and $K'$ ($2880$\,s),
centering the field on the elliptical galaxy lensing the $z=4.04$ arcs.
The sky background was $H_{\rm sky}=13.85$\,mag\,arcsec$^{-2}$ and
$K^{\prime}_{\rm sky}=14.08$\,mag\,arcsec$^{-2}$. The high background
in any near-infrared observations necessitates short individual
exposures to avoid saturation of the detector wells.  We adopted
exposure times of $5$\,s ($H$) and $10$\,s ($K'$), co-adding $12$ such
reads before writing out each data frame.  Hence, for each frame we have
an integration of $60$\,s ($H$) and $120$\,s ($K'$).  Between frames we
offset the telescope, dithering in a $3\times3$ pattern of step size
8\arcsec. This produces an incoherent sequence of disregistered frames,
ensuring that each pixel samples genuine sky frequently. There were 38
independent pointings in $H$ and 24 in $K'$. The NIRC field of view was
oriented at position angle $142^{\circ}$ east of north in order to
avoid dithering the arcs onto brighter objects while retaining a
suitable star in the fixed offset guider.

Photometric calibration was achieved through observations of the UKIRT
Faint Standards FS3, FS27, FS29 \& FS32 (Casali \& Hawarden 1992) over a
range of airmasses, from which we determined the extinction coefficients
($0.06$\,mag\,airmass$^{-1}$ and $0.12$\,mag\,airmass$^{-1}$ for the $H$-
and $K'$-bands, respectively). The $K'$ magnitudes can be transformed to
the standard $K$ filter by using a color term derived from the
standard star photometry:
\[
K=K'- 0.136(H-K).
\]

\subsection{Data Reduction}
\label{subsec:redux}

The near-infrared data reduction involved the removal of two principal
components: an additive term and a multiplicative term. The additive
term is comprised of the bias level and dark current of the detector,
the background contributed by the sky, and the thermal signature of
the telescope. The multiplicative term is the flat-field, dependent on
the intrinsic gain variations of the detector pixels modulated by the
illumination of the array.

As NIRC does not offer a correlated double-sampling readout mode, we
formed our own `super-bias' frames with the same readout pattern as
the science data, medianing many bias frames to reduce the readout
noise.  However, as the bias frames were taken some time away from our
observations of the arcs, they were not appropriate for all pixels,
particularly `warm' pixels with variable dark current.  Those pixels
exhibiting large variability from frame-to-frame were subsequently
identified and masked.  However, other pixels had a time-dependent dark
current which varied on much longer timescales, and could still be used
with a background subtraction determined from the adjacent frames taken
closest in time.

Following dark/bias subtraction, the frames were flat-fielded to remove
the pixel-to-pixel variation in detector gain.  We experimented with
several flat-fielding techniques.  Use of the median of the
disregistered data stack to remove objects in the field, although
producing frames which are cosmetically flat, does not provide a good
representation of the true flat field.  This is because it contains both
the sky background (which is likely to be flat on the $\sim 1$\arcmin\
field of the detector), and the thermal emission of the telescope and
instrument, which will not be constant across the field of view and is
very significant in the thermal-infrared $K$-band.  Hence, this approach
cannot yield a frame that is photometrically flat (\ie, adopting this
technique to flat-field, the photometry of an object will be a function
of the position on the array).  We also obtained dome-flats, formed by
differencing lamp-on and lamp-off images of an illuminated dome spot.
By taking the difference, the thermal signature of the telescope and
dome spot should be removed, and all that remains is illumination of the
array multiplied by the detector flat-field.  However, there may be
non-uniform illumination of the dome spot, although this will vary
smoothly with position on the array as the dome spot is very out of
focus.  Our third method of flat-fielding involved taking a sequence of
dithered twilight sky-flats of equal integration times as the sky
intensity varied.  We performed regression fits for the intensity of
each pixel against the median counts in each frame (the sky level to a
first approximation).  Outliers more than 3-$\sigma$ from the fit were
excluded, to eliminate stars and cosmic rays.  The slope of the
regression line yields the normalized gain for each pixel, and the
intercept is the non-flat thermal contribution (more details are given
in Bunker 1996).  This technique will yield the true flat-field provided
the thermal signature is stable on a timescale of minutes -- the
duration of the twilight sky-flat sequence.  These twilight sky-flats,
which were ultimately used in the calibration, differed from the
dome-flats by a smooth gradient of $\approx 4\%$ across the
field-of-view of the detector.  The robustness of our flat-field was
confirmed by standard star photometry at different locations on the
array.  The dispersion in the recorded calibrated counts was less for
the twilight sky flat than either the dome-flat or the median data-flat.

After removing the bias/dark-current and flat-fielding, the images were
then background-subtracted.  As the background level varied on the
timescale of a few tens of minutes, the background for each frame was
determined by several of the temporally adjacent images.  The {\tt
DIMSUM} package\footnote{{\tt DIMSUM} is the Deep Infrared Mosaicing
Software package developed by P.~Eisenhardt, M.~Dickinson,
A.~Stanford \& J.~Ward.} was used to create object masks for each individual
frame.  The ten frames closest in time were then combined, excluding the
masked objects, to create a background image appropriate for that
particular frame.  Each frame had cosmic rays identified through
unsharpened masking, which differentiated the hard edges of these events
from the softer profile of real objects.

We combined all the background-subtracted, flat-fielded frames,
registering them using offsets determined from the centroid of a bright
star close to the arcs of interest.  Each pixel was rebinned onto a
$2\times2$ grid, and the data registered at the half-pixel level.  All
frames were scaled to zero airmass using the extinction term determined
from the standard star photometry. Optimal inverse-variance weighting
was employed, to maximize the signal-to-noise ratio. Known bad pixels
and previously-identified cosmic rays were flagged for each individual
frame and excluded from the co-addition.  To eliminate other cosmic rays
and artifacts which had escaped previous cuts, the two highest and two
lowest intensities measured for each registered pixel were ignored, and
the inverse-variance weighted average of the other 34 frames (in $H$) or
20 frames (in $K'$) was taken.  In the final mosaics, our $1\,\sigma$
sensitivity in a 1\sq\arcsec\ aperture is
$H(1\sigma)=25.46$\,mag\,arcsec$^{-2}$ and
$K(1\sigma)=24.92$\,mag\,arcsec$^{-2}$ for the central region where all
the images overlap and our noise is lowest.  These limits are entirely
consistent with Poisson statistics based on the sky background and the
readnoise of the detector ($30\,e^{-}$).  Hence, our $5\sigma$ detection
limit for a point source is $H(5\sigma)=23.8^{m}$ and
$K(5\sigma)=23.3^{m}$.  The PSF in our final combined $H$ \& $K'$ images
had a FWHM of 0\farcs5.

\subsection{Analysis}
\label{subsec:anal}

As with the WFPC\,2 imaging, the surface brightness profile of the
lensing galaxy was determined by fitting elliptical isophotes of
increasing radii, and this model was subtracted from the final NIRC
image in each waveband.

The goal of this work is to map out the spatially-resolved color
gradients along the arcs.  The coordinate transform from the
sub-pixelated ($0\farcs075$\,pixel$^{-1}$) NIRC mosaics onto the
intrinsically higher-resolution WFPC2 images was determined using
approximately 40 compact sources common to all frames.  The resultant
pixel scale is $0\farcs1$\,pixel$^{-1}$.  This process, as well as
matching the different orientations and pixel scales, also corrects for
spatial distortions between the two detectors.

The arcs are essentially linear.  We collapse the flux along the short
axis of the arcs in the elliptical-subtracted images, summing over a
fixed aperture width chosen to maximize the signal-to-noise ratio (S/N).
The extracted one-dimensional light profile gives the integrated flux at
each point along the major axis of the arcs.  This is somewhat akin to
the one-dimensional extraction of a two-dimensional spectrum, with the
`dispersion' dimension corresponding to the linear distance along the
arcs.

We trained the aperture traces on the F814W image, in which the arcs had
the highest S/N.  We repeated the photometry on corresponding aperture
centers in the other wavebands.  For the HST imaging, where the arcs are
marginally resolved in the transverse direction (FWHM$\approx0\farcs3$),
our extraction width was 5\,pixels ($0\farcs5$),
$\approx1.5\times$\,FWHM, which for a Gaussian profile maximizes the S/N.
For the ground-based near-infrared images, the seeing in the final
mosaics was $\approx0\farcs5$ and the arcs are unresolved in the
transverse direction, so we set our extraction width to be $0\farcs8$.
Fig.~\ref{fig10} shows the intensity of the arcs in each waveband in
terms of linear magnitudes, the flux per unit length along the arc. A
surface brightness `measurement' would be inappropriate for these
linear, unresolved (or barely-resolved) systems.

\placefigure{fig10}

For each image, our aperture width for the photometry is fixed, with the
trace following the centroid of the arcs.  We convert our fixed-width
aperture magnitudes to total linear magnitudes using aperture
corrections for the fraction of the flux estimated to fall beyond our
extraction widths.  The cross-section of a bright star in our field was
block-replicated to emulate the linear nature of the arcs.  Analysis of
the curve-of-growth of the profile perpendicular to the major axis led
us to estimate the aperture corrections to be $-0.05^{m}$ in the HST
images and $-0.25^{m}$ in the NIRC images for the extraction widths
adopted.

To allow a valid comparison between the optical and near-infrared data
with their different intrinsic spatial resolutions, we boxcar-smooth the
intensity maps by summing the flux in the one-dimensional extraction
over the width of the resolution element determined from the
lower-resolution ground-based seeing ($0\farcs5$). Hence, the derived
colors (from our linear magnitudes, using fixed-width apertures) are now
unaffected by differences in the resolution of the optical and
near-infrared imaging. The maps of intensity as a function of distance
along the arcs are plotted in Fig.~\ref{fig10}.

\section{Discussion}
\label{sec:disc}

\subsection{Age and Reddening from Broad-Band Colors}
\label{subsec:genintro}

We wish to compare our spatially-resolved optical/near-infrared
photometry with stellar spectral synthesis models of various
metallicities.  We also consider the effects of extinction by dust
intrinsic to the source and the absorption by foreground neutral gas
(the line and continuum opacity of intervening absorbers).  Lensing is
achromatic, so the colors are not influenced by the fact that the system
is magnified (once the different resolutions of the various images have
been accounted for).

At $z=4.04$, the $(H-K')$ color straddles the 4000\,\AA\,$+$\,Balmer
break.  The amplitude of the break is affected by the age, the initial
mass function, the dust reddening, and the metallicity.  It is most
sensitive to age (Fig.~\ref{fig2}) and relatively insensitive to
dust, because of the small rest-wavelength baseline.  We can break the
age/dust degeneracy to some extent through use of all colors, which span
a significant range of wavelengths.  However, the age/metallicity
degeneracy persists.

Using ages derived from $(H-K')$, we can then use the
optical/near-infrared colors as an excellent diagnostic of dust
absorption.  The F814W filter is centered at a rest wavelength of
$\lambda_{\rm rest}\approx 1600$\,\AA, and the pass-band lies longward
of Ly-$\alpha$.  Hence it is very susceptible to intrinsic reddening by
dust, but is immune to the effects of line blanketing by the intervening
Ly-$\alpha$ forest.

The interpretation of the $(V_{555}-I_{814})$ rest-ultraviolet color is
more complicated since the intrinsic stellar SED is extremely sensitive
to dust extinction at these short rest-wavelengths. In addition, most of
the F555W filter lies shortward of Ly-$\alpha$ from the arcs. Hence, the
F555W flux may have a line contribution from Ly-$\alpha$ (in emission or
absorption), and the continuum will be severely depressed by $z<4$
intervening Ly-$\alpha$ clouds (Madau 1995).

\subsection{Color Gradients Along the Arcs}
\label{subsec:gradients}

\begin{figure}[ht]
\plotone{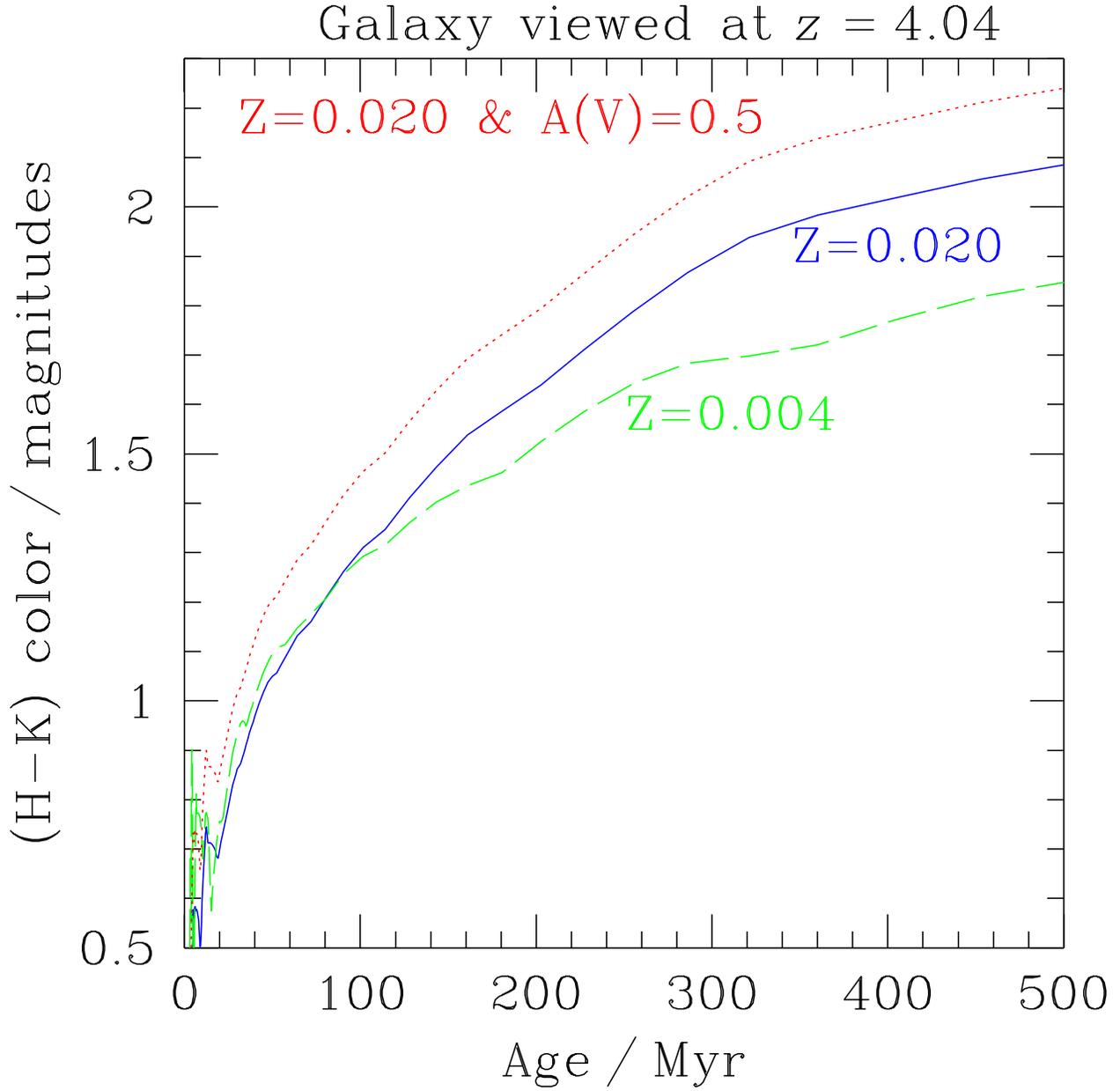}
\figcaption[fig2.eps]{The evolution of the $(H-K)$ color of a galaxy at
$z=4.04$ as a function of the time elapsed since an instantaneous burst
of star formation. The solid curve is the Bruzual \& Charlot model for
Solar metallicity ($Z=0.020$), with the dashed line showing lower
metallicity, $\frac{1}{5}$ solar ($Z=0.004$).  For this redshift, the
$(H-K)$ color is an excellent tracer of the time elapsed since the end
of star formation. The dotted curve is the solar-metallicity model with
a Calzetti-law dust reddening of $A_{V}=0.5^{m}$. \label{fig2}}
\end{figure}

\begin{figure}[ht]
\plotone{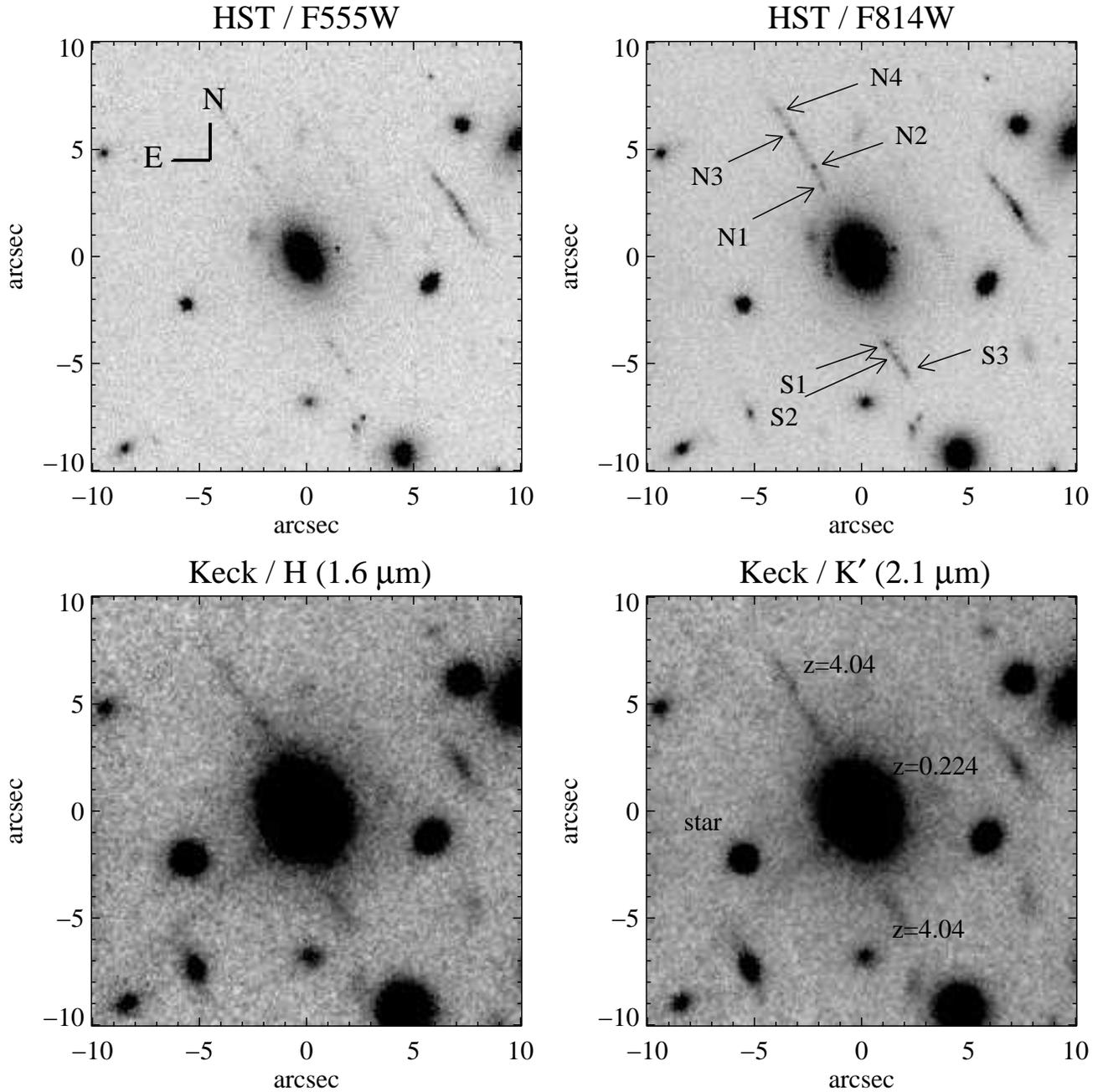}
\figcaption[fig3.eps]{The top panels show archival HST/WFPC\,2 imaging
   of the cluster Abell 2390.  The $z=4.04$ galaxy is the arclet at
   PA=$+23^{\circ}$ that is bisected by the elliptical.  Top left is the
   HST $V$-band (F555W, 8400\,s) which encompasses Ly-$\alpha$, with the
   HST $I$-band (F814W, 10500\,s) top right. The knots which are bright
   in the rest-UV (and so are presumably sites of recent star formation)
   are indicated.  Our Keck/NIRC images were obtained in good seeing
   ($0.4-0.5''$ FWHM) and are shown lower left ($H$, 2280\,s) and lower
   right ($K'$, 2880\,s). The radial feature extending from the center
   of the elliptical to the bottom left corner of the NIRC images is a
   ``bleed trail'' (actually an amplifier hysterisis effect)
   lying along the detector rows; we selected the PA of
   our observations so that this detector artifact did not affect the
   photometry of the $z=4$ arcs. \label{fig3}}
\end{figure}

\begin{figure}[ht]
\plotone{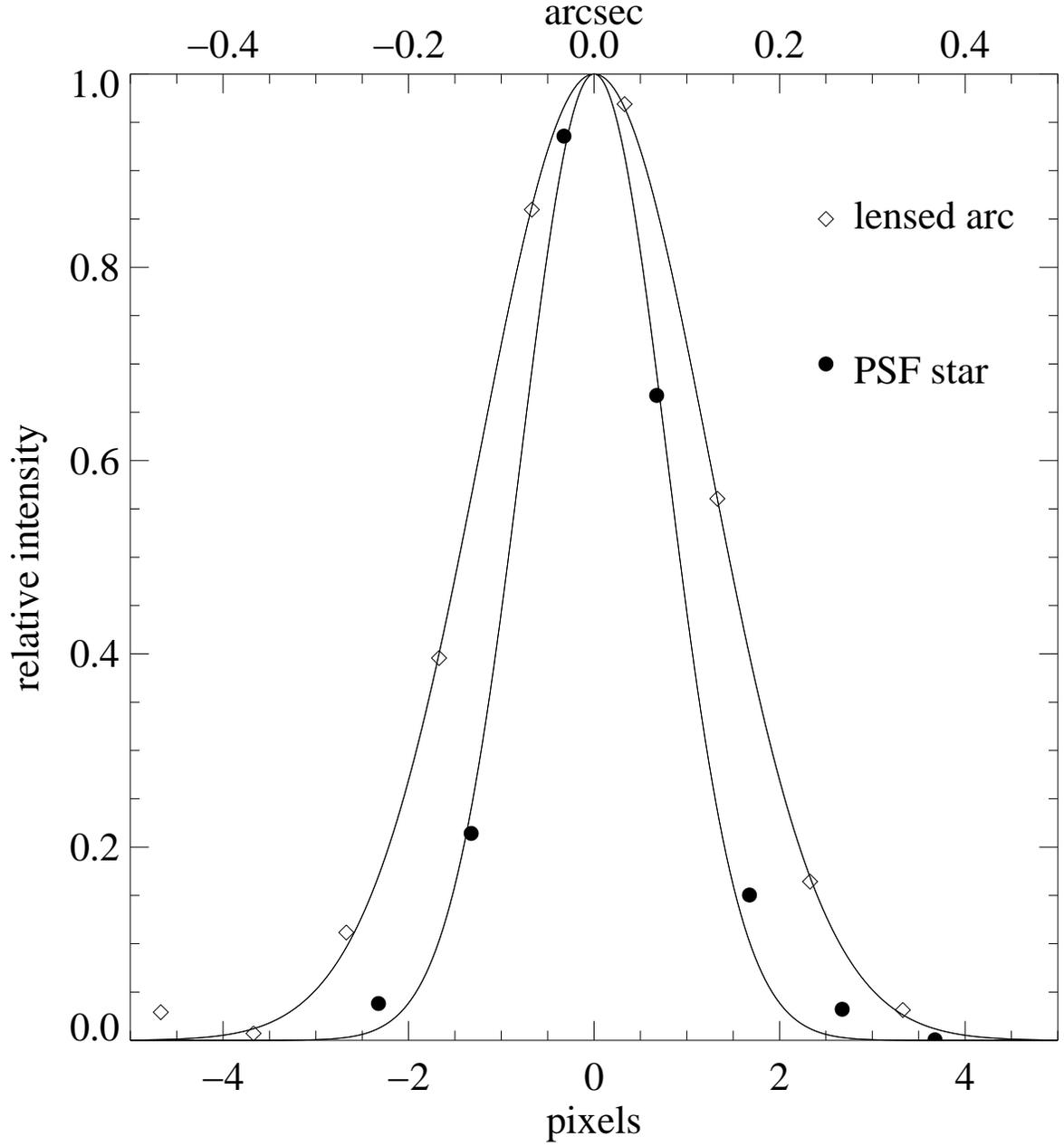}
\figcaption[fig4.eps]{The transverse extent of the $z=4.04$ arcs
(diamond symbols), averaging 0\farcs6 along the northern arc around
N3. A Gaussian fit to the distribution is shown, with
$\sigma=0\farcs123$ (FWHM=0\farcs3). Overplotted is an
appropriately-scaled cross-cut of the nearby bright star ($\sigma\approx
0\farcs078$, FWHM=0\farcs18) which is undersampled by the WFPC\,2 pixels
(0\farcs1). Therefore, the arcs are marginally resolved in their
width. \label{fig4}}
\end{figure}

\begin{figure}[ht]
\plotone{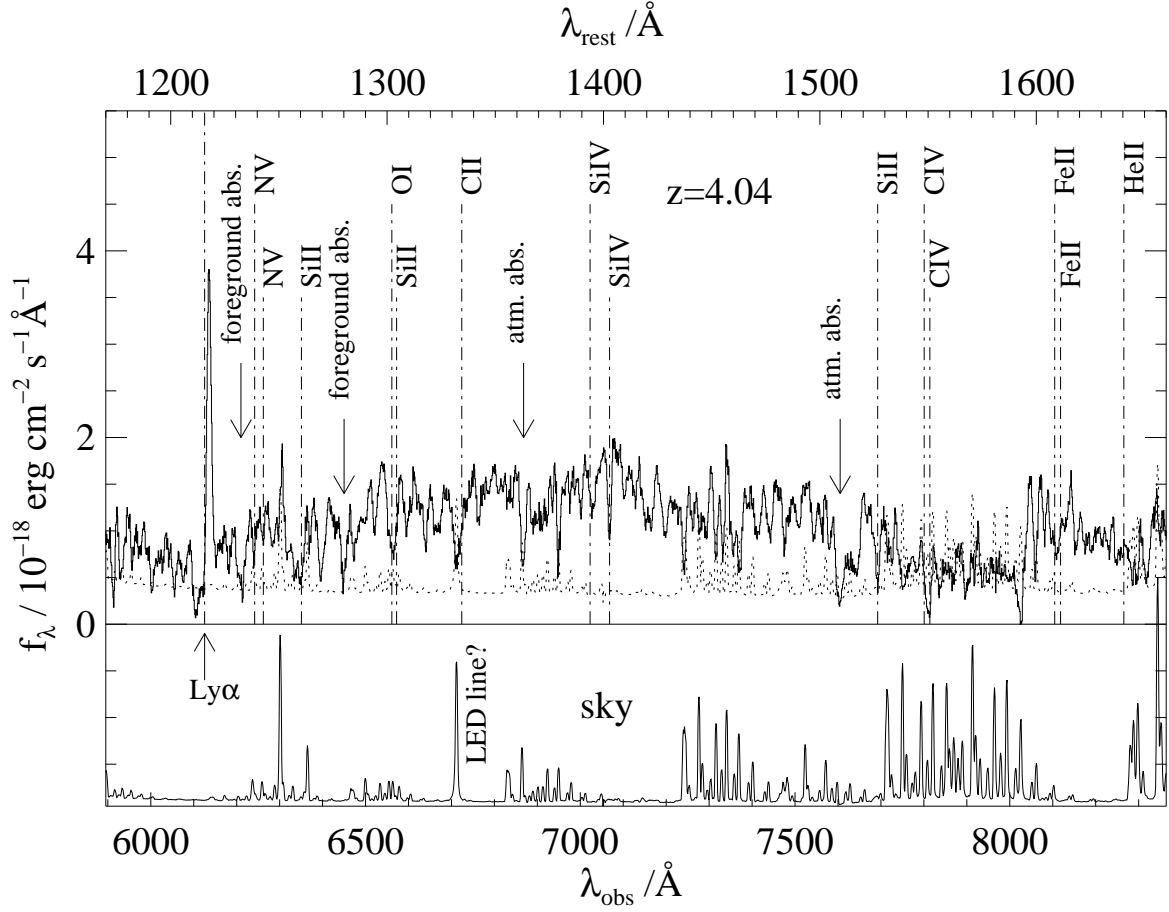} \figcaption[fig5.eps]{One-dimensional
spectral extraction of the continuum in the southern arc, with the
wavelengths of common rest-UV absorption features marked. The extraction
width was 8\,pixels spatially (1\farcs7) and the spectrum has been
smooth with a box-car of 7\,pixels (9\,\AA ). The dotted line shows the
$1\,\sigma$ noise level, and the lower panel displays the sky
spectrum. The prominent emission line in the background spectrum at
6711\,\AA\ does not correspond to a known sky line, and is probably
attributable to an LED light somewhere in the detector's light
path.\label{fig5}}
\end{figure}

\begin{figure}[ht]
\plotone{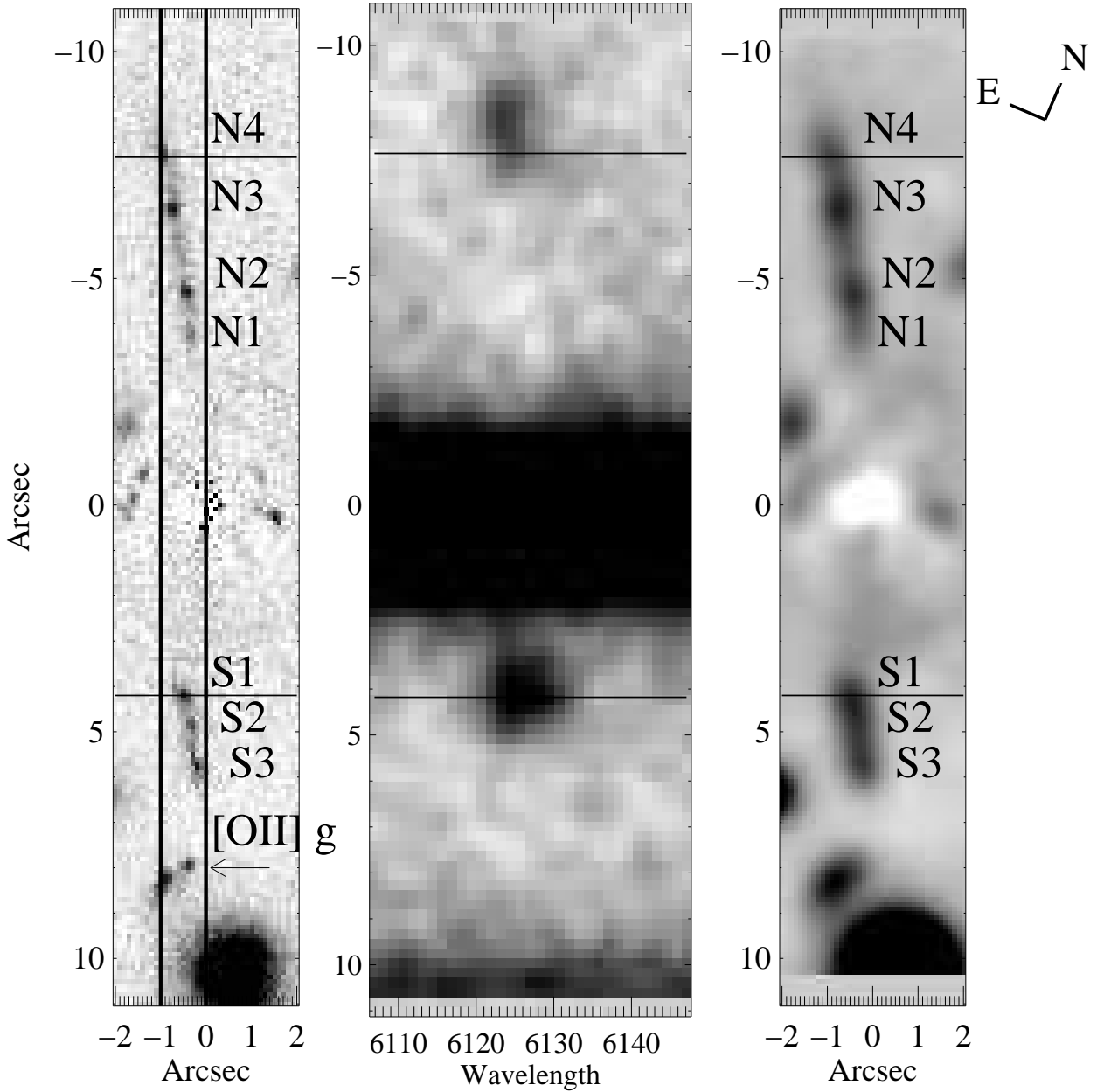} \figcaption[fig6.eps]{Left: the F814W
image with elliptical galaxy model subtracted (note the counter arcs
perpendicular to the axis of the main arcs, predicted by the lens model
of FB98). The area covered by the long-slit optical spectroscopy is
shown (slit axis is vertical). The right panel is this
elliptical-subtracted image, smoothed to $\approx 0\farcs6$ seeing. The
apparent slight over-subtraction of the elliptical at the very center
arises from the finite-pixel sampling skewing the central isophotes; it
does not affect the photometry at the location of the arcs.  Center: the
LRIS spectrum, with the long-slit aligned along the arcs. The dispersion
axis is horizontal, with wavelength increasing from left to right. The
image has been smoothed by convolving with a Gaussian kernel of
$\sigma=1$\,pixel. The positions of N4 ($-7\farcs6$) and S1 ($4\farcs3$)
are indicated by the horizontal bars. Note the spatial range of
Ly-$\alpha$ emission, which extends well beyond the detectable continuum
of the arcs. The knot N4 (top) lies on the edge of the slit, hence the
slight spatial offset of the Ly-$\alpha$ line centroid and the lower
flux compared to the line from S1 (bottom). Also indicated on the left
panel is the $z=1.129$ [\ion{O}{2}]\,$\lambda$\,3727\,\AA\ galaxy also
falling on our spectroscopic long-slit (see
Fig.~\ref{fig7}). \label{fig6}}
\end{figure}

\begin{figure}[ht]
\resizebox{0.75\textwidth}{!}{\includegraphics{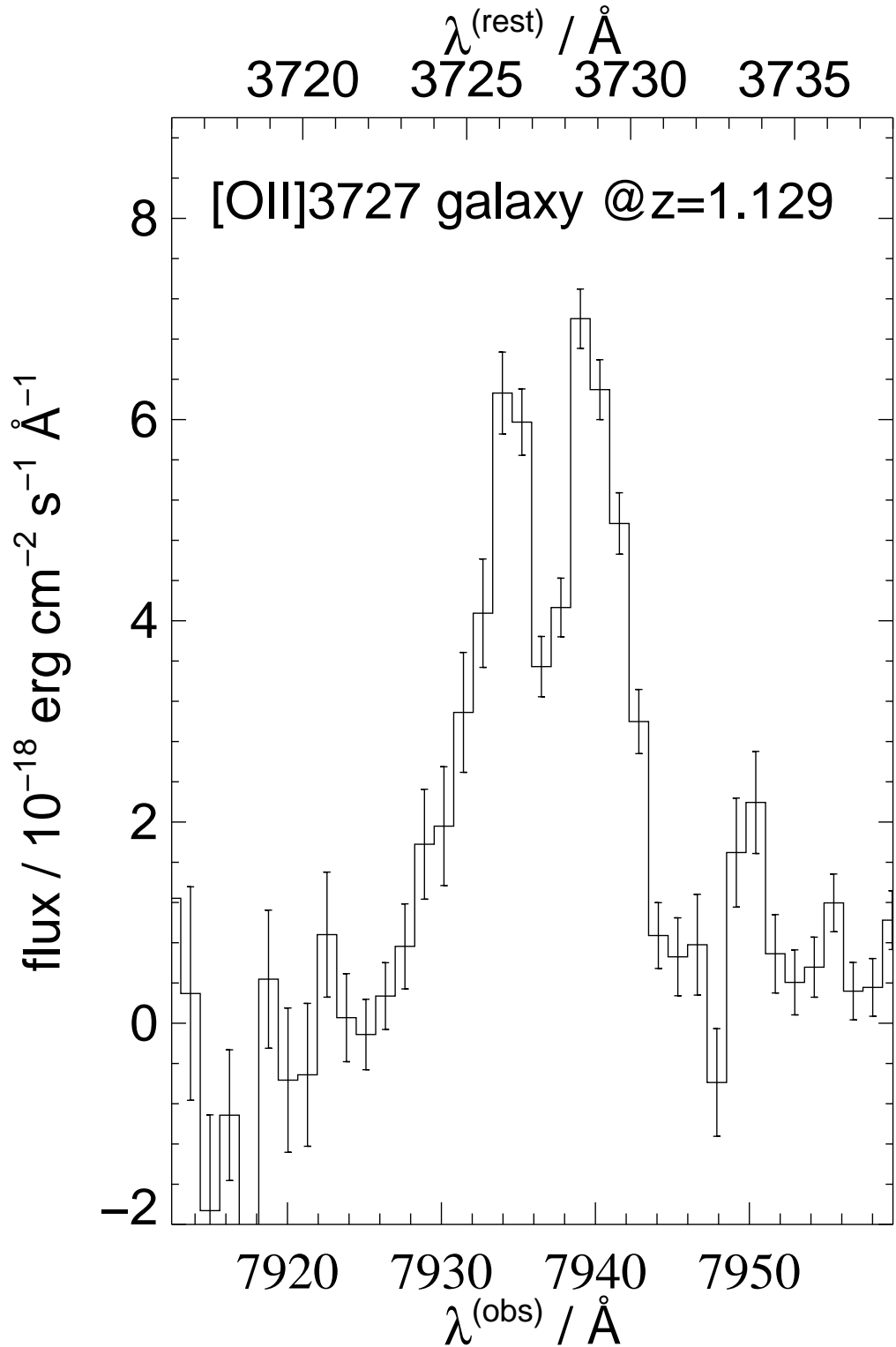}}
\figcaption[fig7.eps]{One-dimensional spectral extraction of the
[\ion{O}{2}]\,$\lambda\lambda$\,3726.1,3728.9\,\AA\ galaxy at
$z=1.1293$, 3\arcsec\ below the southern arc and intercepted by our
spectroscopic long-slit (Fig.~\ref{fig6}). The extraction width is
7\,pixels (1\farcs5). Note that the emission-line
doublet is clearly resolved. We do not see this structure in the
emission lines from the arcs, implying that their origin is not
[\ion{O}{2}]\,3727\,\AA\ at $z=0.64$. \label{fig7}}
\end{figure}

\begin{figure}[ht]
\resizebox{0.75\textwidth}{!}{\includegraphics{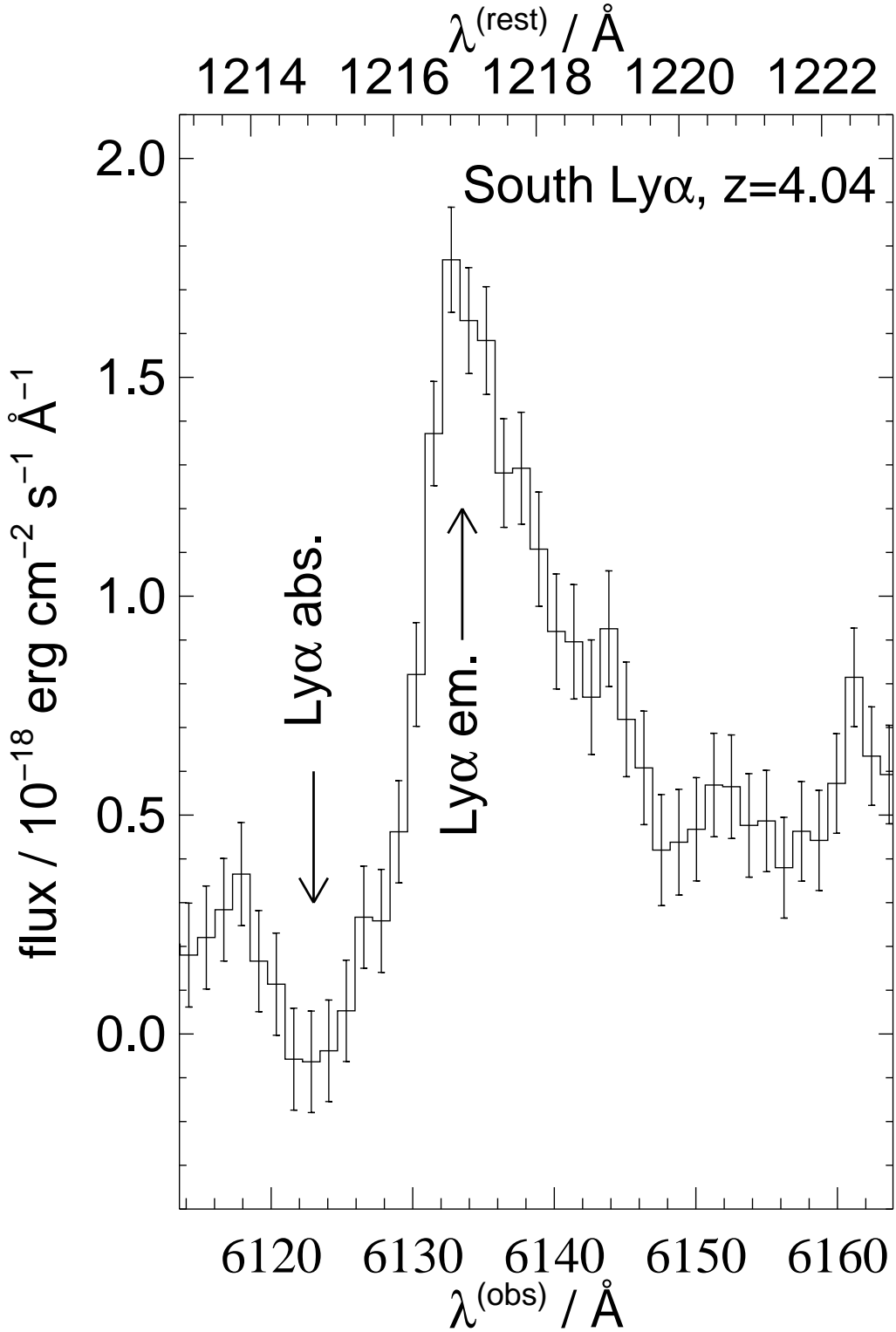}}
\figcaption[fig8.eps]{One-dimensional spectral extraction of the
southern arc, showing the region around Ly-$\alpha$ at $z=4.04$. The
extraction width is 8\,pixels (1\farcs7), and encompasses both line- and
continuum-emission regions. The asymmetric emission line profile is
readily apparent, with the sharp decline on the blue side due to
absorption by \ion{H}{1} within the galaxy -- a blueshifted Ly-$\alpha$
absorption trough is visible from the outflowing \ion{H}{1}).
\label{fig8}}
\end{figure}

\begin{figure}[ht]
\plotone{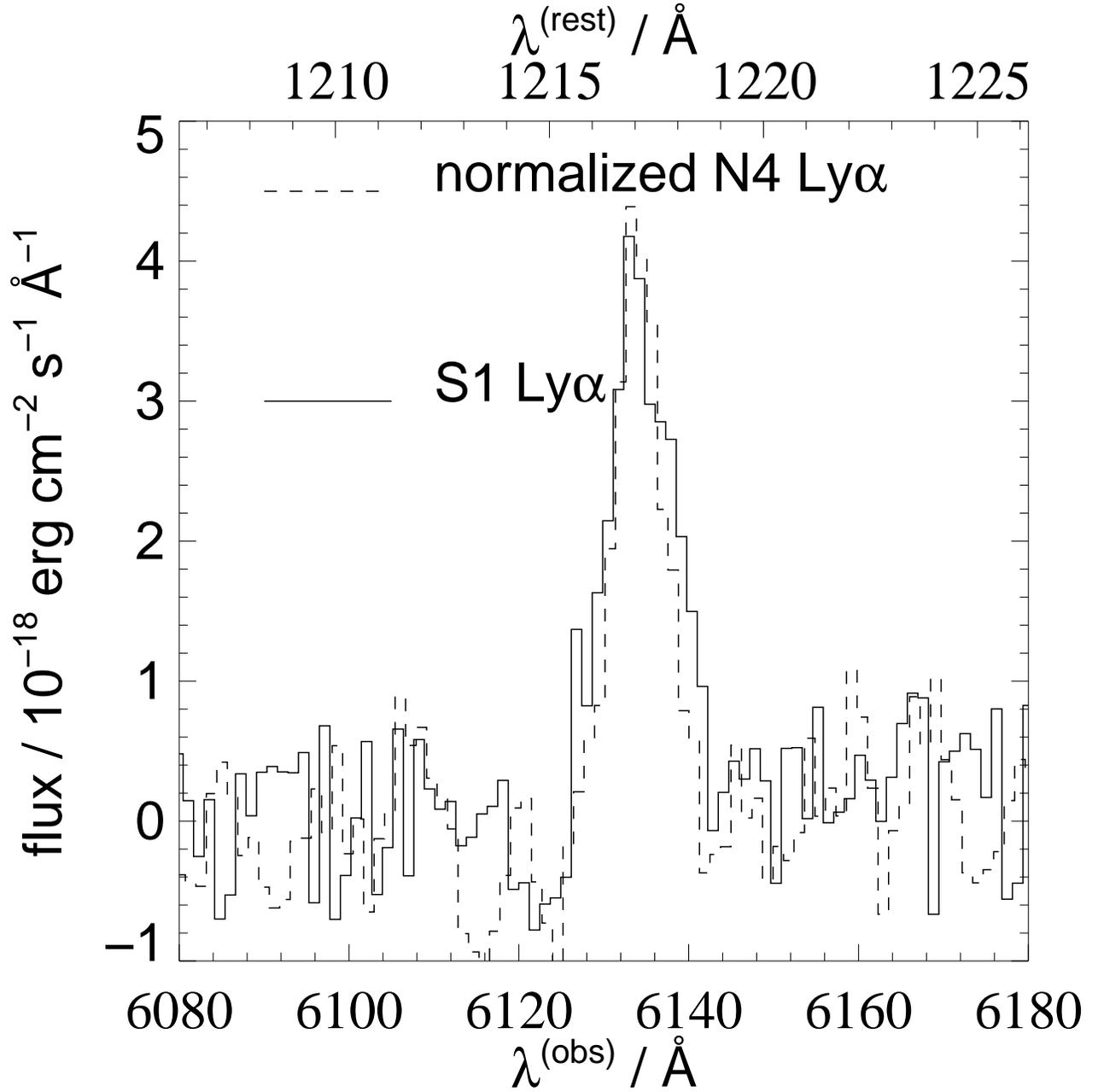}
\figcaption[fig9.eps]{The line profiles of Ly-$\alpha$ emission from the
star-forming knots N4 \& S1 in the northern and southern arcs. The
centroids are identical to within our measurement error ($\Delta
v<50\,{\rm km\,s}^{-1}$), supporting the hypothesis that these are lensed
images of the same source. For comparison purposes, the Ly-$\alpha$ flux
from N4 (which was close to the edge of the spectroscopic long slit) is
normalized to the peak of the S1 emission, which is well-centered. The
integrated line flux from S1 is $f=3.7\times 10^{-17}\,{\rm
ergs\,cm^{-2}\,s^{-1}}$. \label{fig9}}
\end{figure}

\begin{figure}[ht]
\plotone{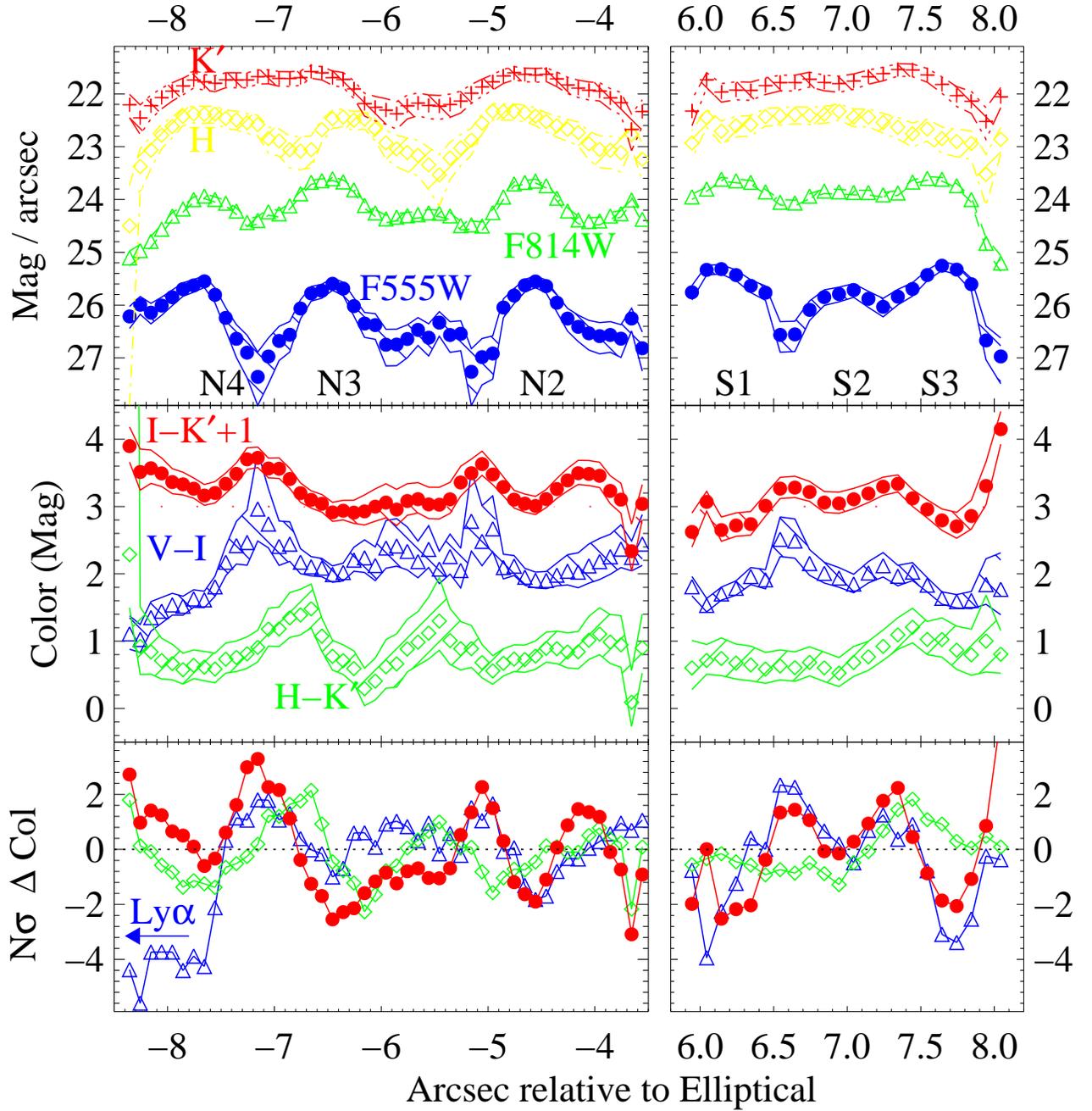} \figcaption[fig10.eps]{Magnitudes (top
panel) and colors (middle panel) of the arcs.  The northern arc is
displayed left, with the southern arc on the right with the same
relative scale.  The $(I_{814}-K')$ color has been offset by $+1^{m}$
for clarity. Linear magnitudes are shown (the brightness per unit
length), and the errors bars correspond to the $1\,\sigma$ Poissonian
noise when averaging over a resolution element ($\approx
0\farcs5$). Note the similarity in the flux-per-unit length in the
northern and southern arcs -- these are probably images of the same
source, and surface brightness is conserved in lensing. The bottom panel
shows the statistical significance of these color gradients: for each
resolution element, the number of standard deviations away from the mean
intergrated color of the arc is displayed. The symbols are the same as
the middle panel -- $(H-K')$ open diamonds, $(I-K')$ filled circles and
$(V_{555}-I_{814})$ open triangles. The variation in color is
significant at the 3--5\,$\sigma$ level. The large deviation in
$(V_{555}-I_{814})$ from the average at $\lesssim -7\farcs7$ is because
of line contamination from the site of Ly-$\alpha$
emission. \label{fig10}}
\end{figure}

We detect the arcs in our near-infrared imaging over the entire spatial
range observed in the WFPC\,2 data.  Fig.~\ref{fig10} shows
spatially-resolved colors along the tangential extent of both major
arcs.  There are statistically significant color gradients at the
3--5\,$\sigma$ level (the lower panels of Fig.~\ref{fig10}).  For the
cases of the prominent knots seen in the WFPC\,2 imaging of the
rest-ultraviolet, the colors are comparatively blue (relatively flat in
$f_{\nu}$ below $\lambda_{\rm rest}\approx4000$\,\AA), with
$(H-K')\approx 0.7^{m}$ in the knots.  From Fig.~\ref{fig2}, this is
entirely consistent with an extremely young OB stellar population
($<20-30$\,Myr).  However, there are significant portions of the
interknot regions that are much redder, with $(H-K')$ colors of
$1.4^{m}\pm0.4^{m}$ between N3 and N4 and $1.3^{m}\pm0.3^{m}$ between N2
and N3, with a firm lower bound of $(H-K')>1^{m}$. From Fig.~\ref{fig2},
where ages are plotted against colors for different metallicities, the
time elapsed since the last major episode of star formation in the
absence of dust is $40-200$\,Myr for metallicites of $Z=0.004$, and
$40-150$\,Myr for solar metallicity ($Z=0.020$).  This is completely
insensitive to upper mass cutoffs greater than $10\,M_{\odot}$, as these
massive stars will have left the main sequence.

A color comparison of the two highest S/N bands, the F814W and $K'$,
shows that the $(I_{814}-K')$ color is anti-correlated with the trace of
the F814W flux (Fig.~\ref{fig10}).  This is to be expected since the
F814W ($\lambda_{\rm rest}\approx1600$\,\AA) traces ultraviolet-bright
sites of recent star formation, whereas $(I_{814}-K')$ is reddest in the
older regions, and/or those with the most extreme dust reddening.

In the northern arc, knots N1--N3 are not adjacent to the sites of
Ly-$\alpha$ emission.  However, the uncontaminated continuum slope of
the knots (\eg, between F814W and $H$-band) is relatively flat in
$f_{\nu}$ (Fig.~\ref{fig10}), implying a young stellar population and
suggestive of ongoing star formation. Extrapolating this to shortward of
Ly-$\alpha$, the continuum depression as inferred from the $(V-I)$ color
may be used to make a direct estimate of $D_A$, the absorption due to
intervening cosmological \ion{H}{1} clouds. Outflowing neutral hydrogen
intrinsic to the source may also play a significant r\^{o}le in
absorbing the blue-wing of Ly-$\alpha$ (\S\,\ref{subsec:lineprof}).
Formally, the $D_{A}$ continuum break (Oke \& Korycansky 1982) is
defined as
\begin{equation}
D_{A}=\left( 1 - \frac{f_{\nu}(1050-1170\,{\rm \AA})_{\rm
obs}}{f_{\nu}(1050-1170\,{\rm \AA})_{\rm pred}}\right) 
\end{equation}
(\eg, Schneider, Schmidt \& Gunn 1991; Madau 1995); we use our best-fit
SEDs (including dust reddening) to predict the F555W flux in the absence
of \ion{H}{1} absorption. Comparison with the actual F555W flux yields a
determination of $D_{A}=0.5\pm 0.1$, after correcting for the fraction
of this filter which lies above the redshifted 1216\,\AA\ break.

The $(V_{555}-I_{814})$ colors in the vicinity of knot N4, from which
there is substantial nearby Ly-$\alpha$ emission, are significantly
bluer than for knots N1--N3.  This difference is consistent with the
emission contributing to the F555W flux.  With this assumption, we
estimate the equivalent width of the Ly-$\alpha$ emission {\em at N4} to
be $W_{\rm rest}\approx 30$\,\AA.  However, as noted in
\S\,\ref{subsec:offsetline}, the Ly-$\alpha$ emission extends further
than N4, with much larger equivalent width over these regions ($W_{\rm
rest}\gg 100$\,\AA). The globally-averaged equivalent width is only
$W_{\rm rest}\approx 10$\,\AA\ (\S\,\ref{subsec:resonant}) typical of
the Lyman break galaxies.

\subsection{Stellar Spectral Synthesis Models}
\label{subsec:bc}

We adopt the spectral evolutionary models of Bruzual \& Charlot (1993;
hereafter, BC) as updated in GISSEL96 (Charlot, Worthey \& Bressan
1996).  As noted by many authors (\eg, Spinrad \etal\ 1997) there are
often disparities between different population synthesis models (\eg,
Worthey 1994; BC; Guiderdoni \& Rocca-Volmerange 1987).  However, we are
interested in a wavelength r\'{e}gime where such differences between
spectral synthesis models are least significant -- the rest-optical.

In fitting stellar populations there is a large number of free
parameters (age, metallicity, reddening, star formation history, IMF,
\etc).  As we have only a limited number of independent color
indices, we begin by fixing all but the age and the reddening, and
explore three variations in metallicity and star formation history.

As with the treatment of Sawicki \& Yee (1998) for the $z\approx 3$ HDF
star-forming galaxies (Steidel \etal\ 1996b; Lowenthal \etal\ 1997), we
have constructed grids of BC models with which to fit the observed
broad-band colors.  We have conservatively adopted the Salpeter (1955)
IMF with a large mass range ($0.1-125\,M_{\odot}$), compatible with that
inferred for local starbursts (Stasinska \& Leitherer 1996.)

We consider two metallicities, both solar metal abundance (for
comparison with the older BC models; $Z=0.020$) and lower metal
enrichment ($Z=0.004$), compatible with the values derived for
high-column-density absorption systems at $z>2$ (20\% solar; \cf, Pettini
\etal\ 1997a), which are probably sight-lines through galaxies.

The final ingredient in the matrix of models is dust extinction.
Longward of Ly-$\alpha$ we only consider dust intrinsic to the source
(once Galactic extinction has been removed).  There has been much debate
about what is the most appropriate reddening law for high-redshift
galaxies (\eg, Meurer \etal\ 1997; Dickinson 1998; Pettini \etal\ 1997b,
1998).  We adopt the extinction law of Calzetti (1997a,b), empirically
derived from local starburst galaxies, which may therefore be more
appropriate to the integrated light of dusty star-forming
regions\footnote{The empirical reddening law for star-forming regions as
determined by Calzetti (1997a,b) is parameterized in terms of the color
excess $E(B-V)$ for the stellar continuum of the OB stars. This is
related to the color excess of the ionized gas (which would be inferred
from nebular line emission observations) by $E(B-V)=0.44\,E_{\rm
gas}(B-V)$. The differential reddening between gas and stars is due to
the fact that the youngest stars (which provide the ionizing flux) are
generally dustier than later stars (which contribute to the continuum,
but no longer to the ionization of the gas). The wavelength dependence
of the reddening is
\[k(\lambda)=2.656(-2.156 + 1.509/\lambda - 0.198/\lambda^{2} +
0.011/\lambda^{3}) + 4.88\hspace{1cm} 0.12\,\mu{\rm m}\le\lambda\le
0.63\,\mu{\rm m}
\]
\[
k(\lambda)=[(1.86 - 0.48/\lambda)/\lambda - 0.1]/\lambda +
1.73\hspace{1cm} 0.63\,\mu{\rm m}\le\lambda\le 1.0\,\mu{\rm m}
\]
where the flux attenuation {\em for the stellar continuum} is given by
$F_{\rm obs}(\lambda)=F_0(\lambda)\,10^{-0.4E(B-V)k(\lambda)}$. Note
that the reddening relation, $A(V)=4.88\,E(B-V)=2.15\,E_{\rm gas}(B-V)$,
differs from that for the Galactic law, $A(V)=3.1E(B-V)$.} than those
derived from individual stars, such as the standard LMC, SMC, and
Galactic reddening laws. In particular, the foreground dust screen
models employed with these extinction laws are likely to be inaccurate
as the geometry of the dust and gas in a star forming region will be
more complex.

We have assembled a grid of SEDs, varying dust extinction between
$E(B-V)=0^{m}$ and $E(B-V)=0.5^{m}$ in increments of $\Delta
E(B-V)=0.005^{m}$, and using 144 age steps
from the BC models (from zero age to 2\,Gyr, the upper-limit on the age
of the Universe at $z=4$ for $q_{0}=0.1$). We then use our photometry
for each location along the arcs to perform a maximum-likelihood fit to
the SED at that point. In our fit, we exclude the F555W flux as it is
prone to both line contamination from Ly-$\alpha$ and
continuum-suppression below $\lambda_{\rm rest}=1216$\,\AA.  Using the
magnitudes measured in the other three wavebands, weighted by the
photometric uncertainty, we minimize the absolute $\chi^{2}$ to compute
the age and reddening for the best-fit model. We have $(n-1)$ degrees of
freedom for fitting a particular model, where $n$ is the number of
pass-bands used in the fit ($n=3$).  Using our subsequent determination
of $D_A$ (\S\,\ref{subsec:gradients}), we repeated the analysis using in
addition a F555W flux corrected for intervening absorption, but found
this did not substantively affect the best-fit SEDs.

We have undertaken this analysis using different star formation
histories, and place constraints based on extremes of the Bruzual \&
Charlot models, from single-event (instantaneous) starburst to a
continuous star-formation rate.  Although undoubtedly simplified, these
models bracket all reasonable star formation histories.  Specifically,
we focus on three scenarios:
\begin{itemize}
\item{a metallicity of 20\% solar ($Z=0.004$) and an
instantaneous burst model;} 
\item{solar metallicity ($Z=0.020$) with the
burst model;}
\item{and solar metallicity with a constant star formation
rate.}
\end{itemize}

\subsection{SEDs of Spatially Resolved Stellar Populations}
\label{subsec:seds}

In this sub-section we discuss the fits of model SEDs to the broad-band
photometry of several specific locations with extreme colors.  We use
this to show that non-coeval star formation histories are suggested for
this $z=4$ object, and we also place limits on the degree of dust
obscuration.

\begin{figure}[ht]
\resizebox{0.45\columnwidth}{!}
{\includegraphics{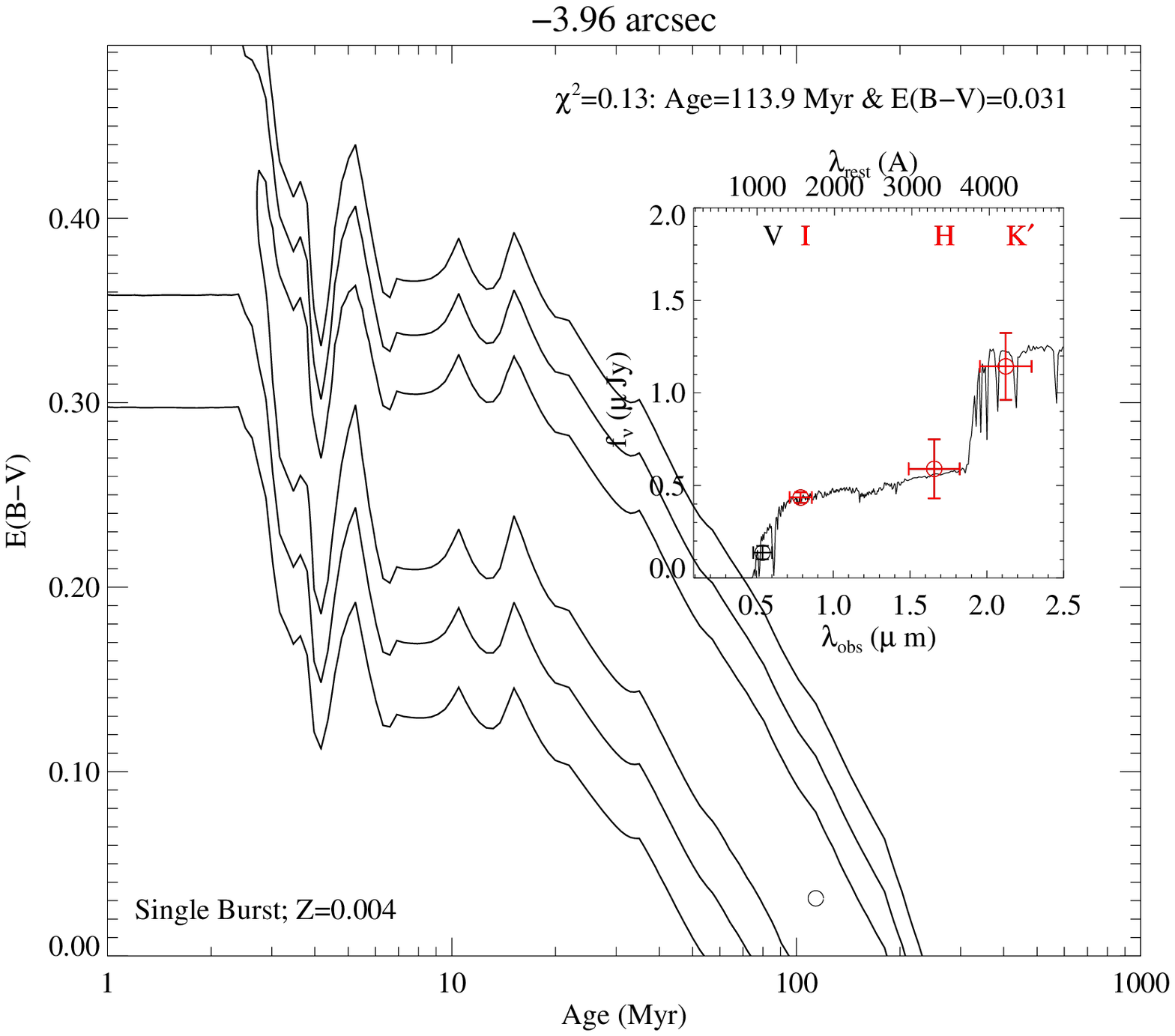}}
\resizebox{0.45\columnwidth}{!}
{\includegraphics{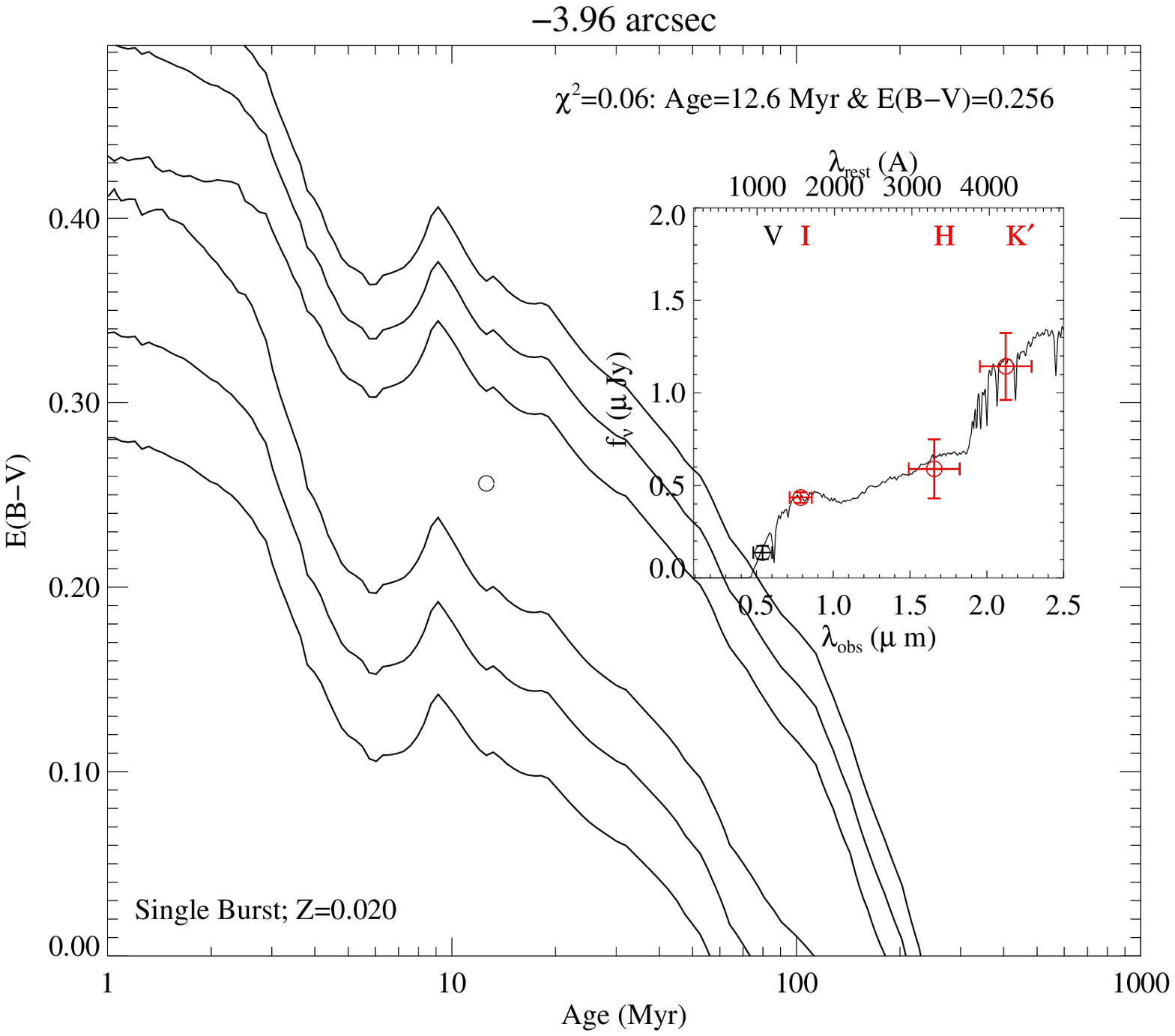}}\hfill
\resizebox{0.45\columnwidth}{!}
{\includegraphics{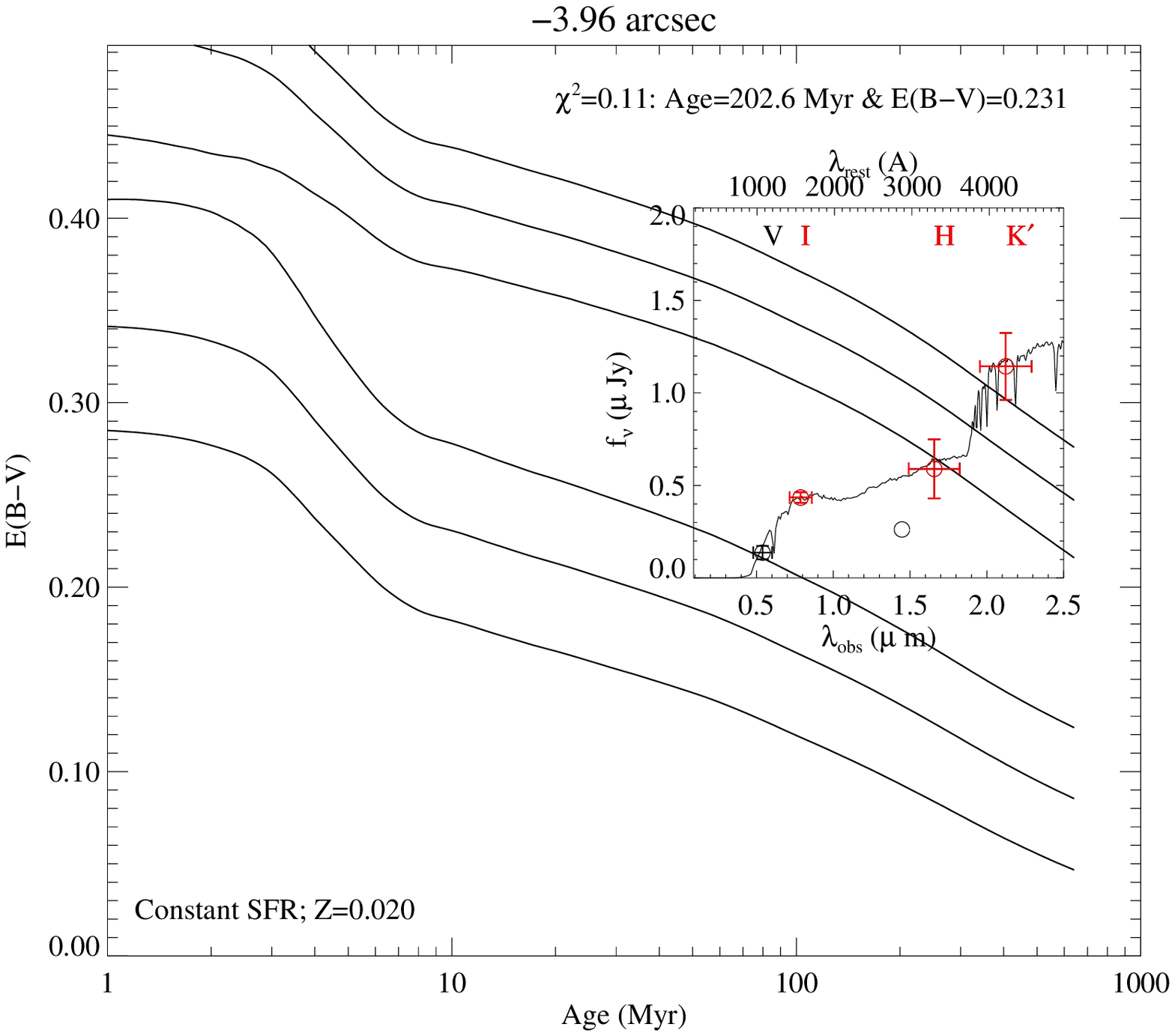}}\hfill
\figcaption[fig11a.eps,fig11b.eps,fig11c.eps]{The parameter space of age
and dust fitting the spectral energy distribution (SED) of a portion of
the linear $z=4.04$ arcs, in this case the interknot region between N1
\& N2. The 1\,$\sigma$, 2\,$\sigma$ and 3\,$\sigma$ contours are shown,
based on the absolute $\chi^{2}$ distribution for $n-1=2$ degrees of
freedom. These correspond to 68\%, 95\% and 99\% confidence
intervals. The formal best fit (lowest $\chi^{2}$) is denoted by a
circle, and the inset panel shows the photometry in the four wavebands
plotted over the appropriately-normalized SED for these parameters. Only
the $F814W$ (`$I$-band'), $H$ \& $K'$ are used in the SED fitting; the
$F555W$ (`$V$-band') lies shortward of Ly-$\alpha$ and is subject to
absorption by intervening \ion{H}{1}. The fits are shown for a
metallicity of 20\% solar ($Z=0.004$) and an instantaneous burst model
(a); solar metallicity ($Z=0.020$) with the burst model (b); and solar
metallicity with a constant star formation rate (c). \label{fig11}}
\end{figure}

First we consider the reddest interknot regions (\eg, Fig.~\ref{fig11},
the interknot region between N1 \& N2 at $-3\farcs96$).  Formally, a
huge range of parameter space is allowed within the $1\sigma$ confidence
interval for all three models we have considered.  The best $\chi^2$ fit
of the single-burst models suggests that the SEDs are dominated by a
$\sim100$\,Myr population, with an extinction of $E(B-V)\approx0.1^m$.
A constant star formation rate model allows for an enormous range of
ages with large extinction due to dust (as much as $A_{V}\gtrsim 2^m$).
Although this model is statistically as viable as the single-burst ones,
a constant star formation rate over more than $100$\,Myr would yield a
detectable far-IR flux, which is not seen by recent ISOCAM observations
-- the $3\,\sigma$ limit at $6.7\,\mu$m of $25\,\mu$Jy per 5\arcsec\
beam (Altieri \etal\ 1999; L\'{e}monon \etal\ 1998) constrains the
reddening to be $E(B-V)<0.7^{m}$. An even better upper limit on the dust
obscuration comes from the SCUBA sub-mm cluster lens survey of Blain
\etal\ (1999). Comparing the star formation rate inferred from the
rest-UV continuum (\S\,\ref{subsec:intrinsic}) to the upper-limit on the
SFR from the SCUBA non-detection\footnote{The $3\,\sigma$ flux limit in
the 15\arcsec\ SCUBA beam (encompassing both arcs) at 850\,$\mu$m of
$\sim 4$\,mJy constrains the luminosity to be $L_{\rm 60\mu m}<3\times
10^{11}\,(\mu/10)^{-1}\,h_{50}^{-2}\,L_{\odot}$ ($<9\times
10^{11}\,(\mu/10)^{-1}\,h_{50}^{-2}\,L_{\odot}$), following the
methodology of Hughes \etal\ (1998). The total star formation rate must
be ${\rm SFR}_{\rm FIR}< 70\,(\mu/10)^{-1}\,h_{50}^{-2}\,M_{\odot}\,{\rm
yr}^{-1}$ ($< 200\,(\mu/10)^{-1}\,h_{50}^{-2}\,M_{\odot}\,{\rm
yr}^{-1}$) and the dust mass $M_{\rm dust}<2\times
10^{7}\,(\mu/10)^{-1}\,h_{50}^{-2}\,M_{\odot}$ ($<5\times
10^{7}\,(\mu/10)^{-1}\,h_{50}^{-2}\,M_{\odot}$) for $\beta=1.5$, $T_{\rm
dust}=50$\,K.}  indicates that the {\em total} obscuration of the galaxy
at 1600\AA\ (the observed $I_{814}$-band) cannot exceed a factor of
$\approx 10-15$, so $E(B-V)<0.25^{m}$ for the Calzetti reddening law.

\placefigure{fig11}

Independently, Pell\'{o} \etal\ (1999) have performed SED fits to the
{\em integrated} flux of the arcs, and also conclude that for this
system, continuous star formation models are less satisfactory than a
burst scenario.  For our more favored model (passive evolution after an
episode of star formation), the lower limit on the age is $\gtrsim
40$\,Myr at $95\%$ confidence if we assume $A_{V}\lesssim 0.5^m$ (for
both metallicities), and $\gtrsim 80$\,Myr in the absence of reddening
(Fig.~\ref{fig14}).  There is significant degeneracy in the models
between age and dust, but general conclusions can be drawn by the direct
comparison of the results for knots and interknot regions within the
context of each model grid.

\begin{figure}[ht]
\resizebox{0.45\columnwidth}{!}
{\includegraphics{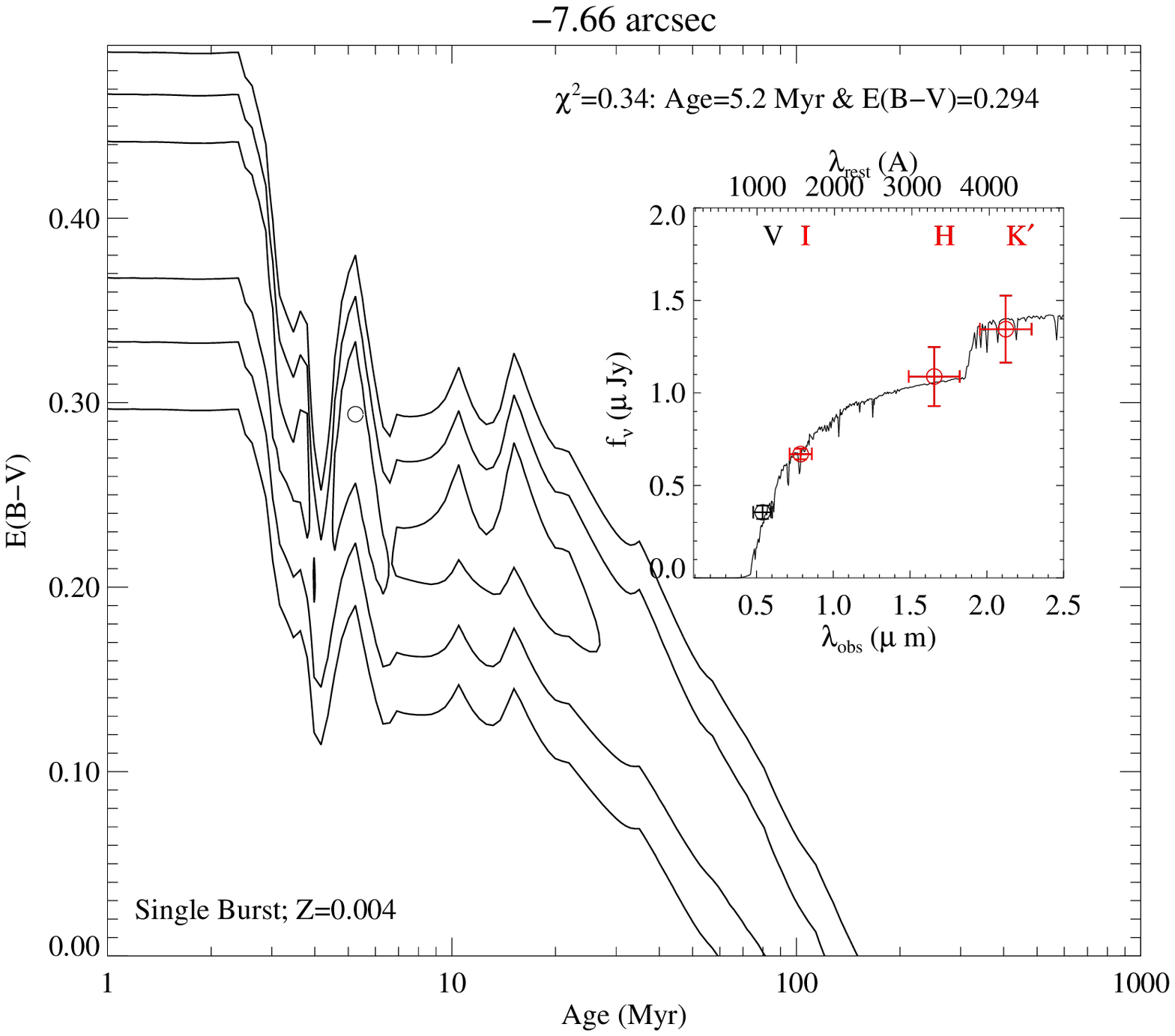}}
\resizebox{0.45\columnwidth}{!}
{\includegraphics{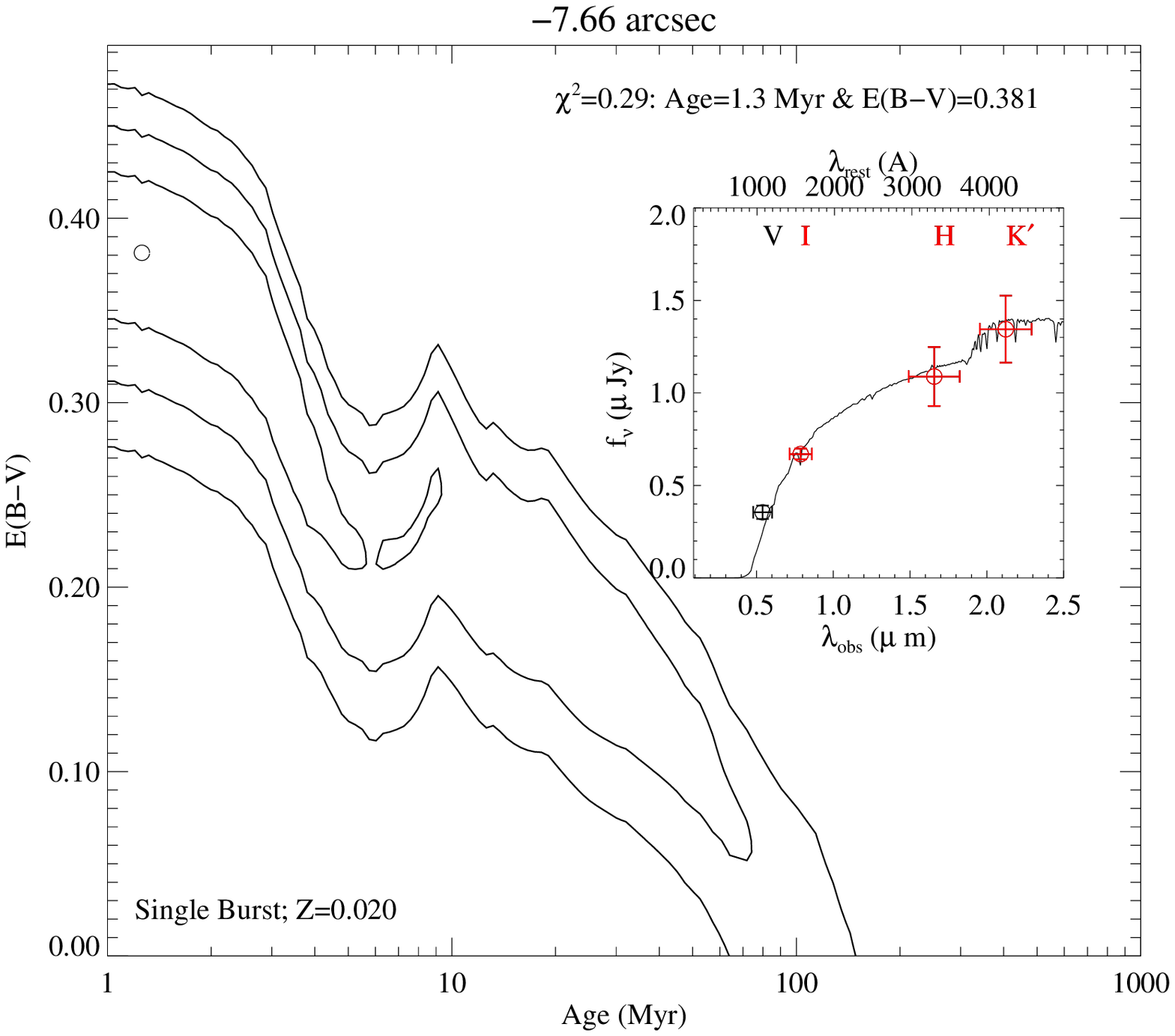}}\hfill
\resizebox{0.45\columnwidth}{!}
{\includegraphics{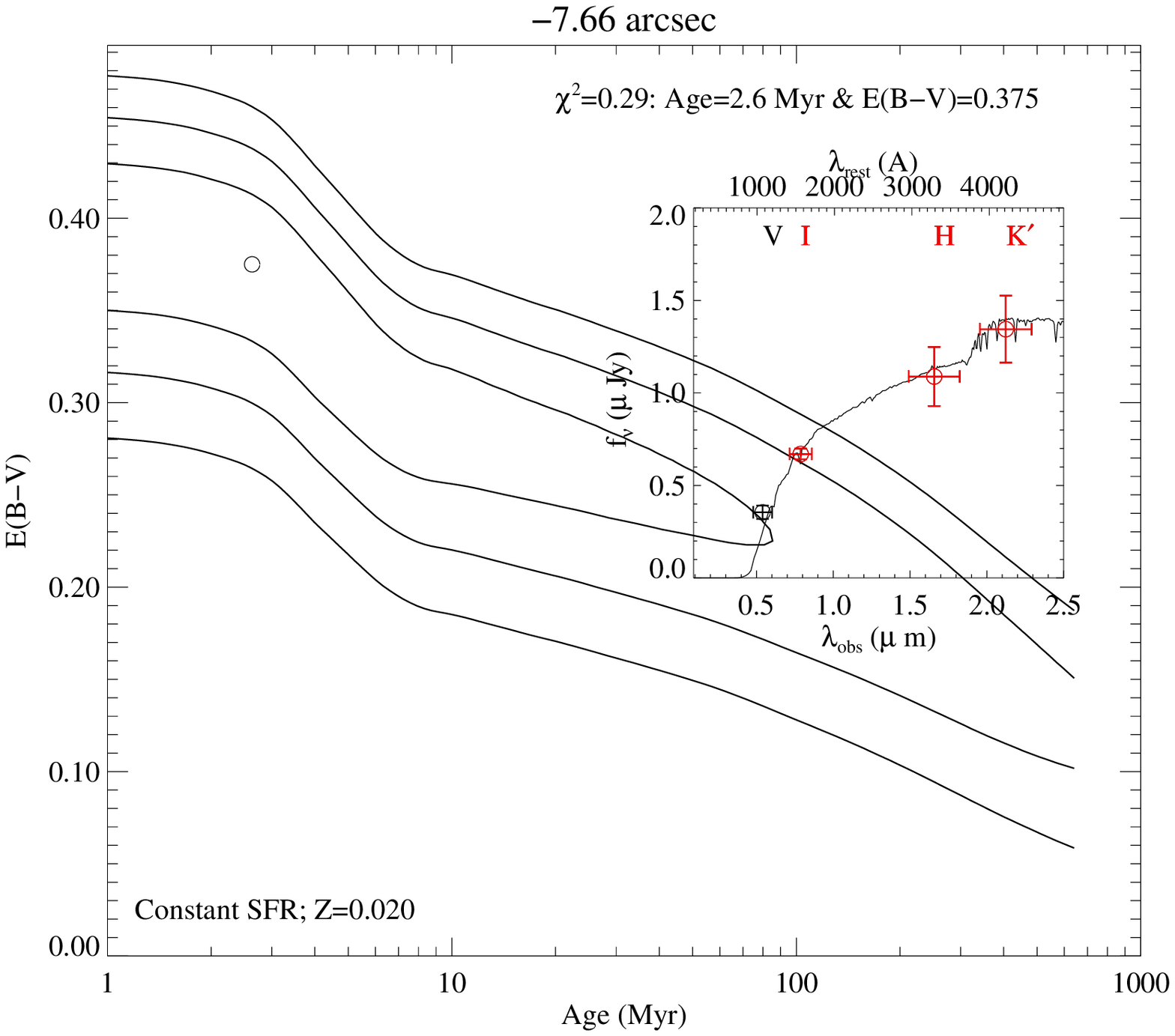}}\hfill
\figcaption[fig12a.eps,fig12b.eps,fig12c.eps]{The age and dust fits to
the SED of knot N4 in the northern arc, which is a site of Ly-$\alpha$
emission. Refer to the caption of Fig.~\ref{fig11} for details. \label{fig12}}
\end{figure}

Comparing our grid of models with the highest surface brightness (knot)
regions yields systematically younger ages than for the interknots.
Clearly, the amplitude of the break between the $H$- and $K'$-bands is
much smaller in the knots (compare the SEDs of Fig.~\ref{fig12} with
Fig.~\ref{fig11}), indicating an intrinsically younger dominant stellar
population -- indeed, the presence of Ly-$\alpha$ in the vicinity of
knot N4 demands this. As we see Ly-$\alpha$ in emission nearby, we know
the upper mass cutoff must exceed $10\, M_{\odot}$ for there to be
sufficient ionizing Lyman-continuum flux, which requires that star
formation has occurred within the last 10\,Myr.  However, in order to
fit the rest-UV optical colors with the Calzetti reddening model, a
higher extinction is preferred in the star-forming knots than in the
older interknot regions.  One could attribute this either to
uncertainties in the true extinction law or, perhaps more physically, to
a dusty environment in these star-forming regions.  In the local
Universe we know that star formation activity is predominantly confined
to molecular clouds with high dust content. Recent sub-mm work on
high-redshift star-forming galaxies (Hughes \etal\ 1998) also suggests a
large extinction in the rest-UV, with re-radiation at thermal
wavelengths by dust perhaps contributing significantly to the diffuse
far-IR background (Schlegel \etal\ 1998; Hauser \etal\ 1998).  Indeed,
just $A_V=0.1^m$ of optical extinction will shift 25\% of the rest-frame
UV flux into the far-IR\footnote{An absolute limit on dust obscuration
in the high-$z$ Universe is provided by the COBE spectrum of the
microwave background, which is a perfect 2.73\,K black-body to within
0.3\% (Mather \etal\ 1994). Significant thermal dust emission from
high-$z$ galaxies would render the spectrum non-Planckian (\eg, Blain \&
Longair 1993), and Mannucci \& Beckwith (1995) show that an average
extinction of only $A_{V}\gtrsim 1^{m}$ would be in conflict with the
COBE measurements (for star formation at $z\gtrsim 5$ and a dust
temperature of $T_{\rm dust}=30$\,K).}.

\placefigure{fig12}

\begin{figure}[ht]
\resizebox{0.45\columnwidth}{!}
{\includegraphics{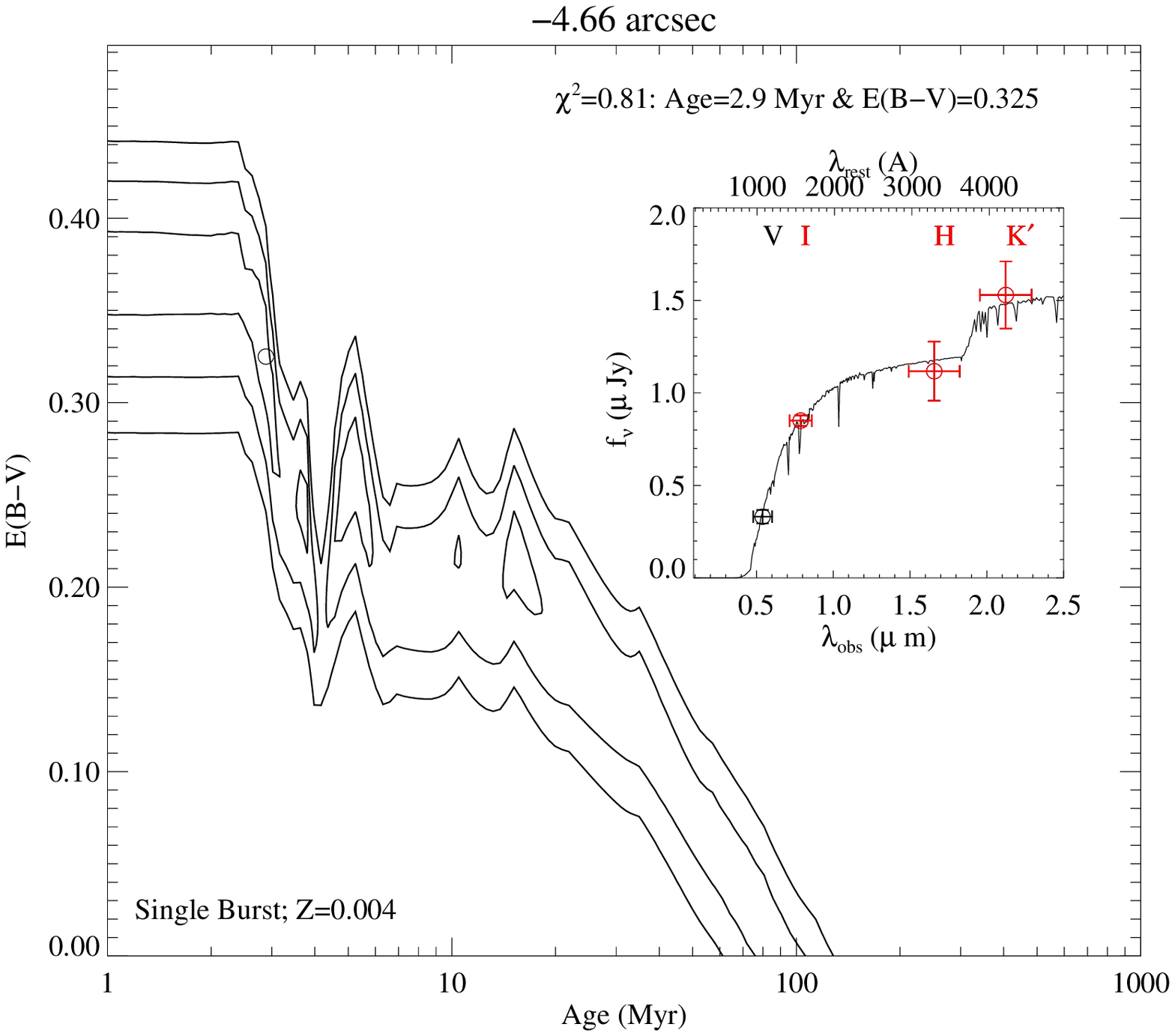}}
\resizebox{0.45\columnwidth}{!}
{\includegraphics{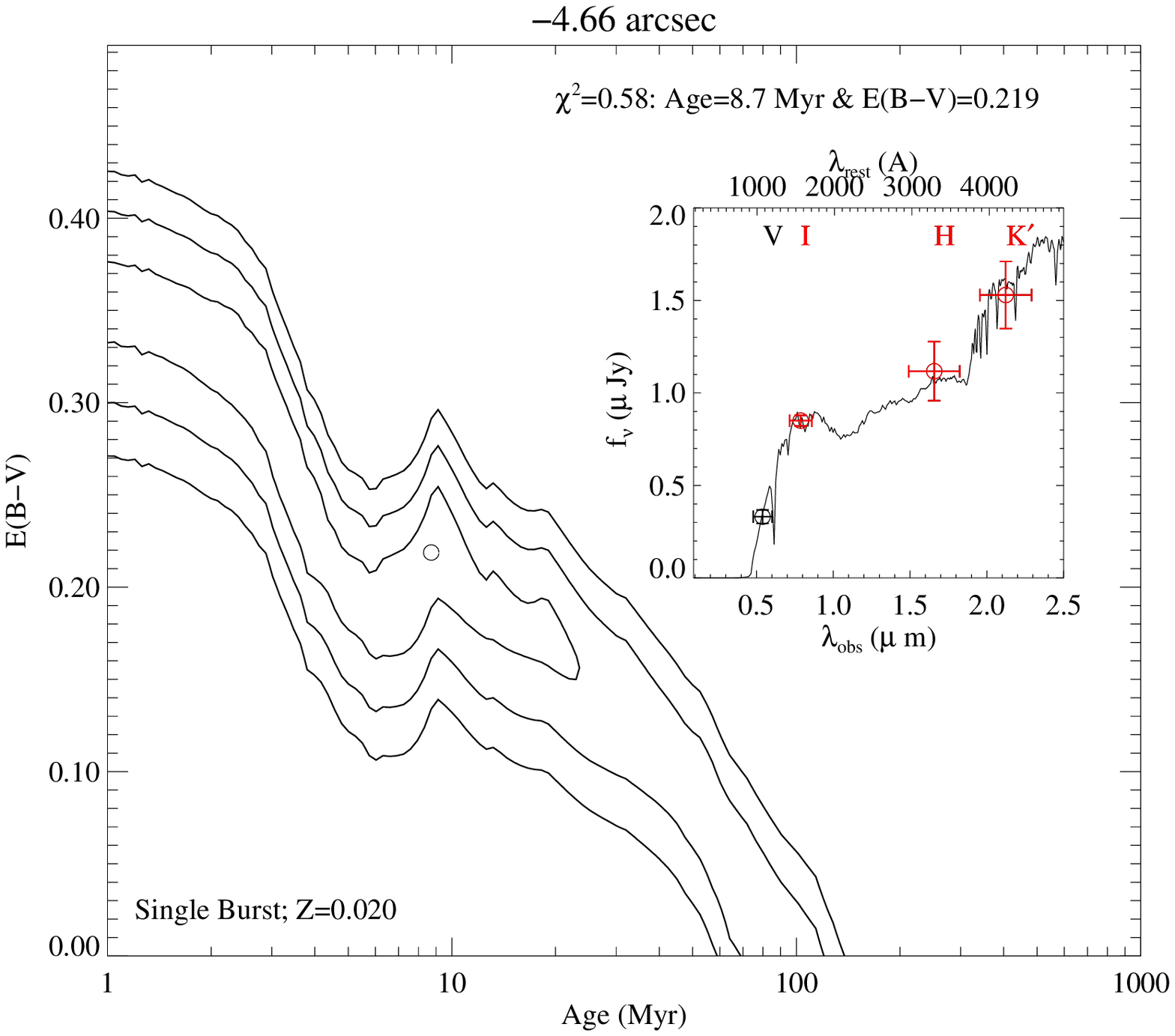}}\hfill
\resizebox{0.45\columnwidth}{!}
{\includegraphics{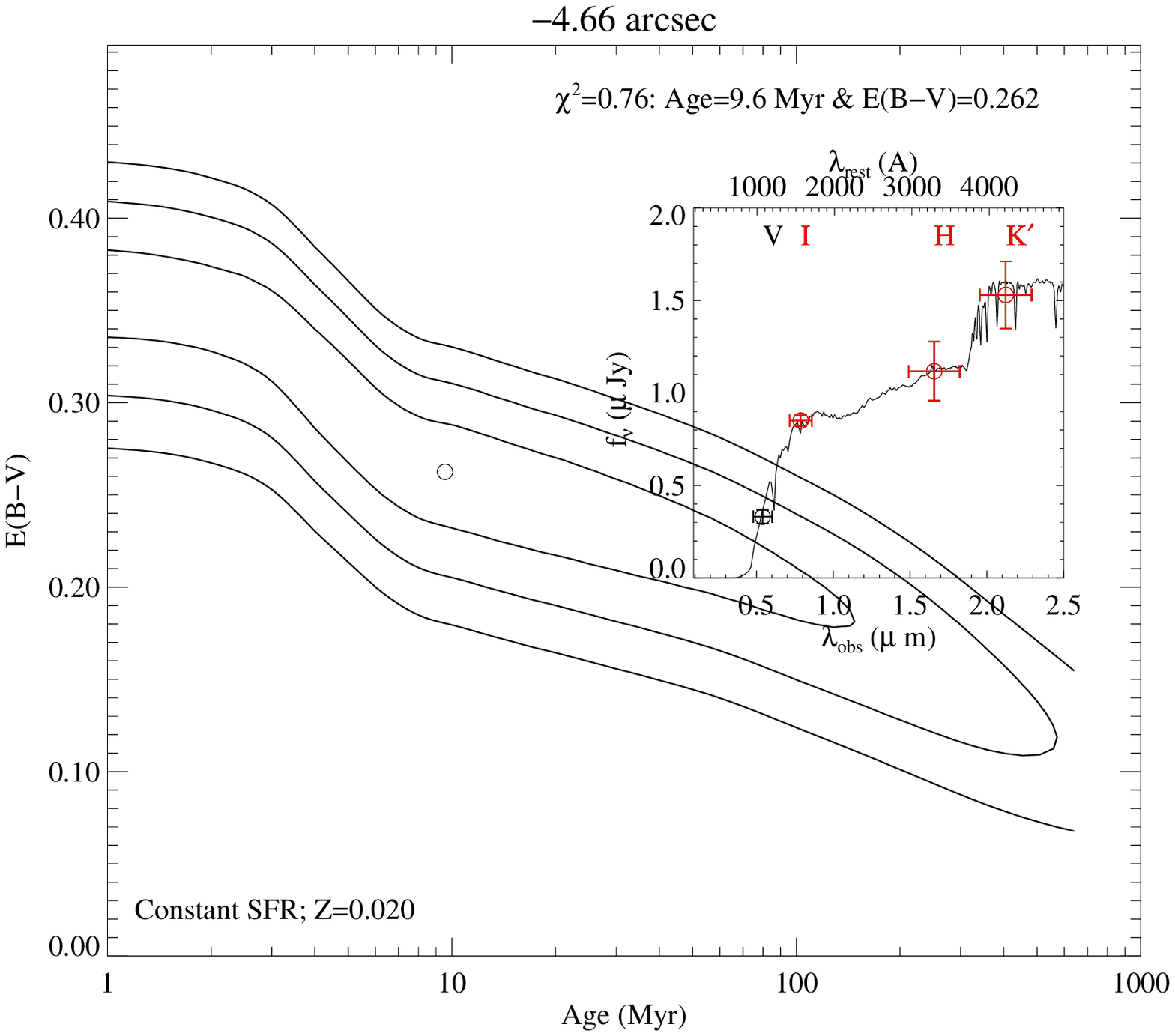}}\hfill
\figcaption[fig13a.eps,fig13b.eps,fig13c.eps]{The age and dust fits to
the SED of knot N2 in the northern arc, which has no detectable
Ly-$\alpha$ emission. Refer to the caption of Fig.~\ref{fig11} for
details. \label{fig13}}
\end{figure}

\begin{figure}[ht]
\plotone{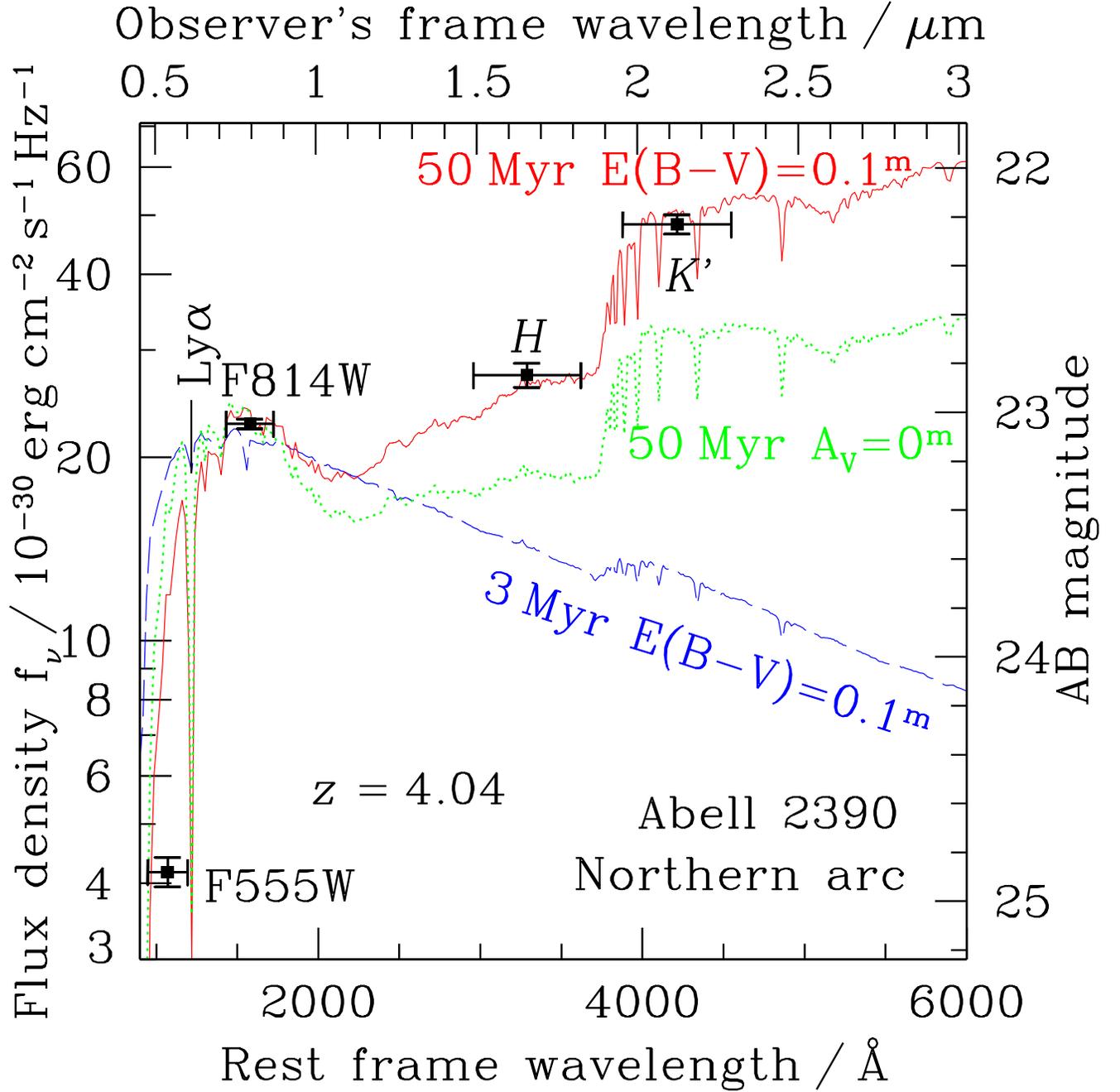} \figcaption[fig14.eps]{The broad-band
optical/near-IR flux from the entire northern arc. Also plotted are
reddened instantaneous-burst stellar population models viewed at various
ages, arbitrarily normalized to the flux measured from the HST/WFPC\,2
F814W image. The flux in F555W is severely attenuated by the opacity of
the intervening Ly-$\alpha$ forest.  The colors are best reproduced by a
stellar population $\sim 50$\,Myr old, with {\em in situ} dust reddening
of $E(B-V)\approx 0.1^{m}$.  Note that at $z=4.04$, the strong
Balmer\,+\,4000\,\AA\ break due to the older stars lies between the $H$-
and $K'$-filters.  The AB magnitude system (right axis) is defined as:
${\rm AB} = -2.5\,\log_{10}f_{\nu} - 48.57$ (Oke \& Gunn 1983) where the
flux density, $f_{\nu}$, is in units of
ergs\,cm$^{-2}$\,s$^{-1}$\,Hz$^{-1}$. \label{fig14}}
\end{figure}

Study of a prominent knot without Ly-$\alpha$ emission (N2) shows that
it is still well-fit by a relatively young stellar population
(Fig.~\ref{fig13}), comparable to that of N4.  Indeed, the absence of
Ly-$\alpha$ emission is not incompatible with an actively star-forming
region; based on local starburst work (Kunth \etal\ 1998) and studies of
the $z\approx 3-4$ Lyman-break population (Steidel \etal\ 1996a,1999),
we know that Ly-$\alpha$ emission is absent in about half the sample of
`normal' galaxies. This is probably attributable to selective extinction
of this resonant line after multiple scatterings
(\S\,\ref{subsec:resonant}), with the large dispersion in the equivalent
widths of Ly-$\alpha$ arising from the complex dependence on the
geometry of the ISM and the internal kinematics.

\placefigure{fig13}

As originally demonstrated by Bunker \etal\ (1998), and recently
confirmed by the work of Pell\'{o} \etal\ (1999), the broad-band colors
(Table~4) integrated along the entire northern arc (Fig.~\ref{fig14})
show that the overall spectrum is best fit by a more evolved ($\sim
50$\,Myr) population with extinction intrinsic to the source of
$A_{V}\sim 0.3^m$.  This implies that there are older stars outside the
actively star-forming \ion{H}{2} regions.  In some areas,
$(H-K)\approx 1.2^m$ (Fig.~\ref{fig10}), indicating that $\sim100$\,Myr
has elapsed since the major episode of star formation (Fig.~\ref{fig2}).


\placefigure{fig14}

It is interesting to note that without the luxury of the
spatially-resolved colors, the integrated flux would be most heavily
dominated by the luminosity and the colors of the recent starburst
regions.  This would lead to the erroneous interpretation that the
object is a prim\ae val galaxy, observed during its first episode of
star formation.  This paper shows that this is demonstrably incorrect.

\subsection{The Origin of the Line Emission}
\label{subsec:resonant}

An unusual aspect is that the Ly-$\alpha$ line emission is not
coincident with the UV-continuum seen in the HST imaging
(\S\,\ref{subsec:offsetline}).  The line emission regions only overlap
partially with the northern-most edges of the two arcs, and extend
$\sim1-2$\arcsec\ beyond.  Therefore, the localized equivalent width for
much of the Ly-$\alpha$--bright region is huge, as the continuum is
undetected. The local $3\,\sigma$ lower-limit on the equivalent width is
$W_{\rm obs}>500$\,\AA\ in the observed frame.  This corresponds to
$W_{\rm rest}>100$\,\AA\ ($3\,\sigma$) in the rest-frame at
$z=4.04$. This lower bound is still consistent with the uppermost-end of
the theoretical Ly-$\alpha$ equivalent widths from stellar synthesis
models of star-forming regions (\eg, Charlot \& Fall 1993) for an
extremely young stellar population ($<10$\,Myr) whose UV flux is
dominated by O-stars. However, the observed Ly-$\alpha$ emission from
star-forming galaxies is invariably much weaker when spatially averaged
over the entire rest-UV extent of the galaxy. The Ly-$\alpha$ emission
is typically $W_{\rm rest}=5-30$\,\AA\ at $z\ga 3$ (\eg, Steidel \etal\
1996a; Warren \& M\o ller 1996), or even in absorption.  Such large
equivalent widths as we see from the emission-line regions of these
lensed arcs are observed in AGN, although the absence of strong
\ion{N}{5}\,1240\,\AA\ \& \ion{C}{4}\,1549\,\AA\ emission from our
spectroscopy (\S\,\ref{subsec:lineprof}) and the lack of X-ray
emission\footnote{The $z=4.04$ arcs lie outside the minimum X-ray
contour in the 28\,ks ROSAT/HRI image of Pierre \etal\ (1996). The
$3\,\sigma$ detection threshold within a 5\arcsec\ aperture (matched to
the ROSAT PSF) is $6\times 10^{-14}\,{\rm ergs\,s^{-1}\,cm^{-2}}$,
adopting the conversion of $7.0\times 10^{-11}\,{\rm
ergs\,s^{-1}\,cm^{-2}\,count^{-1}}$ for $0.2<E<2$\,keV,
$N$(\ion{H}{1})$=6.7\times 10^{20}\,{\rm cm}^{-2}$ and a spectral shape
$\alpha_{X}=1$ (where $f_{\nu}\propto \nu^{-\alpha_{X}}$).  For a
magnification of $\mu$, the $3\,\sigma$ X-ray limit constrains the
source as being $L_{X}<0.5\,({\mu}/{10})^{-1}\,L^{*}_{X}$ on the QSO
luminosity function of Boyle \etal\ (1993), where
$L^{*}_{X}=10^{43.9}\,h_{50}^{-2}\,(1+z)^{2.8}\,{\rm ergs\,s}^{-1}$ for
$q_{0}=0$, determined from QSOs out to $z\approx 3$.}  argue against the
hard UV spectrum of an active nucleus.  The velocity width of
Ly-$\alpha$ is also significantly narrower than seen in the broad-line
regions of AGN, even after correcting for the self-absorbed blue wing
(\S\,\ref{subsec:lineprof}).

We hypothesize that the observed line emission is actually powered by
the neighboring \ion{H}{2} regions. Lyman continuum photons are
generated in the star-forming knot, N4 (and its corresponding southern
image, S1), consistent with the spectral energy distribution (SED) fits
detailed in \S\,\ref{subsec:seds}. Either these Lyman continuum photons
escape along sight-lines relatively free of gas and dust, or (perhaps
more likely) these ionizing photons would be reprocessed by \ion{H}{1}
as Ly-$\alpha$, with their escape paths greatly lengthened through
resonant scattering. Only those Ly-$\alpha$ photons which random-walk to
chemically unenriched regions (\ie, areas where star formation has yet
to occur) will escape quenching by dust. This may favor the diffusion of
Ly-$\alpha$ photons northward of N4 \& S1, away from the body of the
galaxy. Clouds of neutral hydrogen surrounding the galaxy could then act
as external `mirrors' for the Ly-$\alpha$ and any Lyman continuum
photons leaking from the galaxy, and it is from these regions that we
observe the line emission; if the scattering mechanism is primarily
\ion{H}{1}, then the cross-section of the Ly-$\alpha$ photons is $\sim
1000$ times greater than for the UV continuum, explaining why scattered
continuum is not detected. Similar effects have been reported in radio
galaxies (\eg, Villar-Mart\'{\i}n, Binette \& Fosbury 1996) and damped
Ly-$\alpha$ systems (M\o ller \& Warren 1998).

To be more quantitative about this model, we consider the efficiency of
creation and transport of the Lyman continuum photons necessary to power
the observed Ly-$\alpha$ recombination line. We consider the southern
Ly-$\alpha$ emission, as its measured flux is less likely to be subject
to spectroscopic slit losses than the northern region. The adjacent
star-forming continuum knot, S1, has a magnitude of $I\approx 24.8^{m}$,
which is a luminosity density of $L_{\nu}^{\rm 1600\,\AA}=7.8\times
10^{27}\,(\mu/10)^{-1}\,h_{50}^{-2}\,{\rm erg\,s^{-1}\,Hz^{-1}}$
($22\times 10^{28}\,(\mu/10)^{-1}\,h_{50}^{-2}\,{\rm
erg\,s^{-1}\,Hz^{-1}}$). In the absence of extinction of the line, This
would correspond to a star formation rate\footnote{The relation between
the flux density in the rest-UV around $\approx 1500$\,\AA\ and the star
formation rate (${\rm SFR}$ in $M_{\odot}\,{\rm yr}^{-1}$) is given by
$L_{\rm UV}=8\times 10^{27} {\rm SFR}\,{\rm ergs\,s^{-1}\,Hz^{-1}}$ from
Madau, Pozzetti \& Dickinson (1998) for our adopted Salpeter (1955) IMF
with $0.1\,M_{\odot}<M^{*}<125\,M_{\odot}$ (\S\,\ref{subsec:bc}). This
is comparable to the relation derived from the models of Leitherer \&
Heckman (1995).  However, if the Scalo (1986) IMF is used, the inferred
star formation rates are a factor of three higher for a similar
mass-range.} of ${\rm
SFR(S1)}=1.0\,(\mu/10)^{-1}\,h_{50}^{-2}\,M_{\odot}\,{\rm yr}^{-1}$
($2.8\,(\mu/10)^{-1}\,h_{50}^{-2}\,M_{\odot}\,{\rm yr}^{-1}$), and a
production rate of Lyman-continuum photons of $\Psi_{\rm LyC}\approx
1.4\times 10^{53}\,(\mu/10)^{-1}\,h_{50}^{-2}\,{\rm photons\,s}^{-1}$
($4.0\times 10^{53}\,(\mu/10)^{-1}\,h_{50}^{-2}\,{\rm
photons\,s}^{-1}$). However, assuming case~B recombination, the observed
Ly-$\alpha$ line flux (\S\,\ref{subsec:lineprof}) requires only 40\% of
these Lyman continuum photons from S1. This could be provided by an
unobscured light cone centered at S1, directed towards the Ly-$\alpha$
\ion{H}{1} cloud, with an opening angle $\approx 160^{\circ}$ in the
source plane, which is only $\approx 60^{\circ}$ in the image
plane. This light cone would not be directed towards us, otherwise we
would see Ly-$\alpha$ emission coincident with the UV continuum. Of
course, if ionizing photons or resonantly-scattered Ly-$\alpha$ from the
other (further) regions of continuum emission also reach the \ion{H}{1}
cloud where the Ly-$\alpha$ emission is observed, then the efficiency
can be even lower -- the total efficiency in the generation of
Ly-$\alpha$ from the UV continuum of the entire southern arc
($I=23.1^{m}$, Table~4) need only be 9\%. This is comparable to a recent
estimates by Bland-Hawthorn \& Maloney (1999) and Dove, Shull \& Ferrara
(1999) of the fraction of Lyman-continuum photons from OB associations
escaping the \ion{H}{1} disk of the Milky Way. Dove \etal\ argue that
the \ion{H}{2} regions are density bounded in the vertical direction
(perpendicular to the disk), and it is the escape of these photons
through dynamic chimneys and superbubbles which keeps the diffuse,
ionized medium (the Reynolds layer) ionized to scale heights of $\sim
1$\,kpc. It is conceivable that similar mechanisms are at work in the
$z=4$ galaxy.

Using the spatially-integrated flux in the F814W filter extrapolated to
the continuum just above Ly-$\alpha$ (the F555W filter is subject to
Ly-$\alpha$ forest blanketing over much of its bandpass,
\S\,\ref{subsec:genintro}) implies that the globally-averaged equivalent
width of the Ly-$\alpha$ emission is only 50\,\AA\ in the observed
frame. This value of $W_{\rm rest}=10$\,\AA\ is typical of the Lyman
break galaxy population at high redshift (\eg, Steidel \etal\ 1996a).

We now consider the transport of the photons. The radiation must travel
$\approx 1$\arcsec\ beyond the continuum-emission regions of the galaxy
before being scattered by \ion{H}{1} as Ly-$\alpha$ into our
line-of-sight. However, this distance is along the direction of
magnification for these linearly-stretched arcs, so the physical
separation is not large on account of this amplification --
$0.6\,(\mu/10)^{-1}$\,kpc ($1.1\,(\mu/10)^{-1}$\,kpc). We note that
without the luxury of this magnification, with ground-based resolution
we would not be able to resolve the spatially-distinct continuum and
line-emission regions -- it would look like a normal Lyman-break star
forming galaxy, with a typical Ly-$\alpha$ emission line equivalent
width (integrated over the entire galaxy). Even with HST, the sub-kpc
spatial separation between the line- and continuum-emission would be at
the very limit of the WF resolution (0\farcs1) if the galaxy were not
lensed.

\subsection{Intrinsic Properties of the Lensed Galaxy}
\label{subsec:intrinsic}

It has been suggested that there is a limiting surface brightness of
star-forming regions, equivalent to a luminosity surface density of
$2\times 10^{11}\,L_{\odot}\,{\rm kpc}^{-2}$ (Meurer \etal\ 1997). We
consider the knot N4 in the northern lensed arc, which is mostly likely
the site of star formation responsible for the Ly-$\alpha$ emission. We
measure the area occupied by the UV continuum of this young knot, rather
than the area of the Ly-$\alpha$ emission, which is probably scattered
significantly (\S\,\ref{subsec:resonant}). After deconvolution with the
WFPC\,2 point spread function, the solid angle subtended by the
star-forming knot N4 is 0.05\,arcsec$^{2}$ (out to the half-light
radius), corresponding to $0.2\,(\mu/10)^{-1}\,h^{-2}_{50}\,{\rm
kpc}^{2}$ ($0.6\,(\mu/10)^{-1}\,h^{-2}_{50}\,{\rm kpc}^{2}$) for
$q_{0}=0.5$ ($q_{0}=0.1$), and its magnitude is $K'=22.6^{m}$,
equivalent to $M_{B}=-18.9^{m}+2.5\log_{10}(\mu/10)+5\log_{10}h_{50}$
($M_{B}=-20.0^{m}+2.5\log_{10}(\mu/10)+5\log_{10}h_{50}$) or
$L_{B}=5.7\times 10^{9}\,(\mu/10)^{-1}\,h^{-2}_{50}\,L_{\odot}$
($L_{B}=16\times 10^{9}\,(\mu/10)^{-1}\,h^{-2}_{50}\,L_{\odot}$), where
$M_{B\odot}=+5.48$ (Allen 1973).  The mean surface brightness is
therefore $1.1\times 10^{11}\,L_{\odot}\,{\rm kpc}^{-2}$, within the
limiting value suggested by Meurer \etal\ (1997) but significantly less
than that derived for the star-forming knot in the lensed arc G1
($z=4.92$) behind CL1358+62 ($3\times 10^{12}\,L_{\odot}\,{\rm
kpc}^{-2}$; Franx \etal\ 1997).  Our result is independent of the actual
magnification, as the sources are resolved and surface brightness is
conserved, and is also independent of cosmology -- surface brightness is
subject only to $(1+z)^{-4}$ dimming.

Although undoubtedly simplified, lens models (\S\,\ref{subsec:wfpc2};
FB98; Pell\'{o} \etal\ 1999) can be used to recover the intrinsic
properties of the source galaxy.  The inaccuracies in such models are
likely to be factors no more than a few, comparable to the uncertainties
due to the cosmology.  With the estimated linear magnification of
$\mu\sim10$, the physical scale-length of the lensed galaxy is $r\sim
2\,(\mu/10)^{-1}\,h_{50}^{-1}$\,kpc
($3\,(\mu/10)^{-1}\,h_{50}^{-1}$\,kpc), a fairly compact half-light
radius similar to that seen in high-resolution HST imaging of the
$z\approx 3$ Lyman break population (Giavalisco, Steidel \& Macchetto
1996).  The knots N4 \& S1 associated with the Ly-$\alpha$ emission
occupies $\approx$10\% of the arcs, hence we infer a size for the
\ion{H}{2} regions of $\approx 200(\mu/10)^{-1}\,h_{50}^{-1}$\,pc
($\approx 300(\mu/10)^{-1}\,h_{50}^{-1}$\,pc), comparable to that
observed for star-forming \ion{H}{2} regions in the local Universe such
as the starburst knots A \& B in NGC\,1741 (Conti, Leitherer \& Vacca
1995).  The dynamical crossing time for such regions is likely to be
less than $20$\,Myr, so an episode of star formation with comparable
duration (cohesive over these timescales) is not unphysical.  The total
luminosity density of the star-forming knots N1--N4 is about
$L_{\nu}^{\rm 1600\,\AA}\approx 5\times
10^{28}\,(\mu/10)^{-1}\,h_{50}^{-2}\,{\rm ergs\,s^{-1}\,Hz^{-1}}$
($14\times 10^{28}\,(\mu/10)^{-1}\,h_{50}^{-2}\,{\rm
ergs\,s^{-1}\,Hz^{-1}}$).  This is equivalent to about
$10^5\,(\mu/10)^{-1}\,h^{-2}_{50}$ ($3\times
10^5\,(\mu/10)^{-1}\,h^{-2}_{50}$)~O7-stars.  The total star formation
rate inferred from the continuum around $\lambda_{\rm rest}=1600$\,\AA\
is $\approx 5\,(\mu/10)^{-1}\,h^{-2}_{50}\,M_{\odot}$\,yr$^{-1}$
($14\,(\mu/10)^{-1}\,h^{-2}_{50}\,M_{\odot}$\,yr$^{-1}$), adopting the
conversion from Madau, Pozzetti \& Dickinson (1998). This SFR estimate
from the continuum is more reliable and significantly higher than the
value derived from the total Ly-$\alpha$ line flux
(\S\,\ref{subsec:lineprof}), implying that globally this line is
selectively extinguished through resonant scattering by a factor of
$\approx 10$ relative to the UV continuum (\S\,\ref{subsec:resonant}).

Introducing a dust correction to the rest-ultraviolet luminosity
significantly increases the estimated star-formation rate.  From
\S\,\ref{subsec:seds} we infer an intrinsic extinction of
$E(B-V)=0.1^{m}$ ($A_V\approx0.5^m$) which corresponds to an attenuation
in the rest-ultraviolet of a factor of three at 1600\,\AA.  Hence, we
estimate the true star formation rate of this stellar system to be
$\approx 15\,(\mu/10)^{-1}\,h^{-2}_{50}\,M_{\odot}$\,yr$^{-1}$ ($\approx
40\,(\mu/10)^{-1}\,h^{-2}_{50}\,M_{\odot}$\,yr$^{-1}$), which will take
$3\,(\mu/10)$\,Gyr ($1\,(\mu/10)$\,Gyr) to assemble an $L^*$ galaxy with
a mass of $5\times 10^{10}\,h^{-2}_{50}\,M_{\odot}$ in stars, a
timescale equivalent to the Hubble time at $z=4$ and a significant
fraction of the Hubble time at $z=0$. Alternatively, it is likely that
we are witnessing the formation of a sub-$L^{*}$ unit, which will in
time merge with other such units to gradually assemble a more massive
galaxy -- the four bright knots may even be a multiple nucleus arising
from a merger event. Indeed, Pell\'{o} \etal\ (1999) report the
discovery of the lensed image of another galaxy in this field at the
same redshift (but physically distinct) 30\arcsec\ away. This is a
projected distance of only $150\,h^{-1}_{50}$\,kpc
($250\,h^{-1}_{50}$\,kpc) in the source plane, which would be crossed in
$\approx 1$\,Gyr at 100\,km\,s$^{-1}$. Hence, these two $z=4.04$
galaxies may have merged by the current epoch.

The absolute magnitude in the rest-ultraviolet (based on the $I$-band)
is $M^{AB}_{1600}=-20.1^{m}+2.5\log_{10}(\mu/10)+5\log_{10}h_{50}$
($M^{AB}_{1600}=-21.2^{m}+2.5\log_{10}(\mu/10)+5\log_{10}h_{50}$), which
places it at around $0.6\,(\mu/10)^{-1}\,L^{*}_{1600}$ on the luminosity
function determined by Dickinson (1998) for the $z\approx 3$ population
of $U$-band drop-out galaxies, which have a characteristic absolute
rest-UV magnitude of $M^{AB*}_{1600}=-21^{m}$ ($q=0.5$, with no dust
correction).  Recent results from $z\approx 4$ Lyman break galaxies show
an minimal evolution in the luminosity function from $z\approx 3$ to
$z\approx 4$ (Steidel \etal\ 1999).  From the rest $B$-band (our $K'$
image), the absolute magnitude is $M_B\approx
-21.0^{m}+2.5\log_{10}(\mu/10)+5\log_{10}h_{50}$ ($M_B\approx
-22.1^{m}+2.5\log_{10}(\mu/10)+5\log_{10}h_{50}$), which translates to a
luminosity of $3.9\times 10^{10}\,(\mu/10)^{-1}\,h^{-2}_{50}\,L_{\odot}$
($10.8\times 10^{10}\,(\mu/10)^{-1}\,h^{-2}_{50}\,L_{\odot}$). For
$q_{0}=0.5$, this comparable to $L^{*}_{B}$ from surveys of the local
Universe (\eg, $M^{*}_B\approx -21.0^{m}+5\log_{10}h_{50}$ from the APM;
Loveday \etal\ 1992). For $q=0.1$, the $B$-band luminosity of the lensed
galaxy would be slightly above the local $L^{*}_{B}$, perhaps $L\approx
3\,(\mu/10)^{-1}\,L^{*}_{B}$. This almost certainly reflects the young
age of this $z\approx4$ galaxy, rather than a massive system with a mass
in stars exceeding that of an $L^{*}$ galaxy at $z=0$.

\section{Conclusions}
\label{sec:conc}

This paper has described a multi-waveband study of a $z=4$
galaxy. Combining archival HST/WFPC\,2 imaging with deep near-infrared
data taken with Keck/NIRC in good seeing, we have measured the
spatially-resolved colors along a pair of $z=4.04$ arcs, gravitationally
lensed by the rich cluster Abell~2390. Comparison of the optical and
near-IR photometry with a suite of spectral evolutionary models has
enabled us to map spatially the underlying stellar populations and
differential dust extinction. Optical spectroscopy has shown that there
are regions with Ly-$\alpha$ in emission, and that these are adjacent to
some of the bright knots seen in the optical HST images which sample the
rest-frame UV. We have shown that the line emission is likely to be
powered by star formation in these knots, rather than by an active
nucleus. We see the Ly-$\alpha$ line morphology extending significantly
beyond the UV continuum, which we attribute to resonant scattering from
\ion{H}{1}.  In the bright knots, the SEDs are consistent with a very
young stellar population ($<10$\,Myr) or ongoing star
formation. However, our near-IR imaging reveals significantly redder
portions between the prominent knots, and we have shown that for these
locations along the arcs, the most probable SEDs are those viewed
$\gtrsim 100$\,Myr after an episode of star formation, with modest dust
extinction of $E(B-V)\approx 0.1^{m}$. There is degeneracy in the models
between dust reddening and age for the optical/near-infrared colors, but
the most extreme scenario where the color gradients are solely due to
heavy dust reddening of an extremely young stellar population are
strongly ruled out by upper limits in the far-infrared/sub-mm from
ISO/SCUBA. It is therefore unlikely that
this $z=4$ system in a true `prim\ae val' galaxy, viewed during its
first major burst of star formation. Rather, our results suggest that
the star formation history of this system has not been coeval, with
current activity concentrated into small pockets within a larger, older
structure.  The evidence suggests at least two stages of star formation
events are underway. This is not a monolithic collapse at $z\approx 4$
-- the galaxy must have begun forming at $z>4.5$.  Taken in the context
of the recent renaissance in the study of normal galaxies at high
redshift, our results further erode the paradigm of a single ``formation
epoch'' (see also Zepf 1997).

Using the amplification estimated from lens models, we have shown that
the intrinsic properties of the $z=4.04$ galaxy are comparable to the
Lyman-break selected $z\approx 3-4$ population of Steidel \etal\
(1996a,1999).  The current extinction-corrected star formation rate may
be adequate to `build' an $L^{*}$ galaxy over a Hubble time, but a more
likely scenario may be the creation of a sub-unit which will undergo
subsequent merging with nearby systems (such as the other $z=4.04$
galaxy identified by Pell\'{o} \etal\ 1999) to assemble hierarchically
the massive galaxies of today.

\acknowledgments

We are indebted to Brenda Frye and Tom Broadhurst for providing details
of their $z=4$ arcs in advance of publication.  The LRIS data reported
in \S\,\ref{subsec:lris} are used with the kind permission of Hyron
Spinrad, Daniel Stern and Arjun Dey, and we gratefully acknowledge them
for many useful discussions and for their help with the optical
spectroscopy.  Our thanks to Sabine Airieau for assistance in obtaining
the near-infrared imaging, and to Mark Lacy, Eric Gawiser and Jeffrey
Newman for invaluable comments on this manuscript. Daniela Calzetti
graciously provided us with details of the starburst reddening model,
and we have made use of the spectral evolutionary models of Gustavo
Bruzual and St\'{e}phane Charlot. We are grateful to the anonymous
referee for carefully reading the manuscript and for constructive
comments on this work. Some of the data presented herein were obtained
at the W. M.\ Keck Observatory, which is operated as a scientific
partnership among the California Institute of Technology, the University
of California and the National Aeronautics and Space Administration.
The Observatory was made possible by the generous financial support of
the W. M.\ Keck Foundation.  We received excellent assistance while
observing at Keck, and we are grateful to Randy Campbell, Bob Goodrich,
Fred Chaffee and Chuck Sorenson. This work also makes use of images from
the NASA/ESA Hubble Space Telescope, obtained from the data archive at
STScI which is operated by the Association of Universities for Research
in Astronomy, Inc., under the NASA contract NAS\,5-26555.  We are
grateful to Adam Stanford for help with DIMSUM, and to Steve Warren for
many enlightening discussions on the reduction of near-infrared
data. The routine for the rejection of cosmic rays in the optical
spectra was kindly provided by Mark Dickinson.  AJB gratefully
acknowledges a NICMOS postdoctoral fellowship (under grant NAG\,5-3043),
and LAM the support of STScI Grant \#GO-07460.01-96A. MD acknowledges
support from NSF Grant \#AST\,95-28340.

\clearpage



\begin{deluxetable}{ccc}
\tablewidth{33pc} \tablecaption{The coordinates of the $I\approx 18^{m}$
lensing elliptical in Abell~2390 bisecting the $z=4.04$ system of arcs,
and the locations of the various `knots' seen in the continuum images of
the northern and southern arcs. Also tabulated are the coordinates of
the offset star used in the spectroscopy, the location of a $z=1.129$
[\ion{O}{2}]\,$\lambda$\,3727\,\AA\ emission-line galaxy which was
covered in our spectroscopy, and the components of the nearby ``straight
arc.''
\label{tab:coords}}
\tablehead{\colhead{object} & \colhead{$\alpha_{\rm J2000}$} &
\colhead{$\delta_{\rm J2000}$}} 
\startdata
lensing elliptical & $21^{h}\,53^{m}\,33\fs52$ &
$+17^{\circ}\,41^{'}\,57\farcs4$ \nl
offset star & $21^{h}\,53^{m}\,33\fs90$  & 
$+17^{\circ}\,41^{'}\,55\farcs2$ \nl
N1 & $21^{h}\,53^{m}\,33\fs64$ & $+17^{\circ}\,42^{'}\,00\farcs6$ \nl
N2 & $21^{h}\,53^{m}\,33\fs67$ & $+17^{\circ}\,42^{'}\,01\farcs5$ \nl
N3 & $21^{h}\,53^{m}\,33\fs74$ & $+17^{\circ}\,42^{'}\,03\farcs0$ \nl
N4 & $21^{h}\,53^{m}\,33\fs78$ & $+17^{\circ}\,42^{'}\,04\farcs0$ \nl
Ly-$\alpha$ (N)\tablenotemark{1} & $21^{h}\,53^{m}\,33\fs81$ &
$+17^{\circ}\,42^{'}\,05\farcs0$ \nl
Ly-$\alpha$ (S)\tablenotemark{2} & $21^{h}\,53^{m}\,33\fs45$ &
$+17^{\circ}\,41^{'}\,53\farcs9$ \nl
S1 & $21^{h}\,53^{m}\,33\fs43$ & $+17^{\circ}\,41^{'}\,53\farcs3$ \nl
S2 & $21^{h}\,53^{m}\,33\fs41$ & $+17^{\circ}\,41^{'}\,52\farcs8$ \nl
S3 & $21^{h}\,53^{m}\,33\fs38$ & $+17^{\circ}\,41^{'}\,52\farcs1$ \nl
\ion{O}{2}\,3727\,\AA\ emission galaxy\tablenotemark{3} &
$21^{h}\,53^{m}\,33\fs32$ &
$+17^{\circ}\,41^{'}\,49\farcs9$ \nl
straight arc A & $21^{h}\,53^{m}\,34\fs28$  &
$+17^{\circ}\,41^{'}\,54\farcs7$ \nl
straight arc B & $21^{h}\,53^{m}\,34\fs41$  &
$+17^{\circ}\,41^{'}\,58\farcs2$ \nl
straight arc C & $21^{h}\,53^{m}\,34\fs59$  &
$+17^{\circ}\,42^{'}\,03\farcs2$ \nl
\enddata
\tablenotetext{1,2}{Exact coordinates for the line emission are
uncertain as it is undetected in the imaging.
Resolved along the spectroscopic slit axis of PA $23^{\circ}$, the
northern line emission is  $3\farcs7\pm0\farcs2$ and the southern
$8\farcs7\pm0\farcs2$ from the centroid of the elliptical.}
\tablenotetext{3}{This galaxy has two components visible in the WFPC\,2
imaging; the component closest to the slit center is given; the other
lies 0\farcs4 east, 0\farcs5 south.}
\end{deluxetable}

\clearpage

%

\begin{deluxetable}{ccc}
\tablewidth{33pc} \tablecaption{Absorption lines in the continuum of the
arcs. The mean redshift is $z=4.039\pm 0.04$ excluding the Ly-$\alpha$ line.
\label{tab:abslines}}
\tablehead{\colhead{$\lambda_{\rm obs}$\,/\,\AA} &
\colhead{ID, $\lambda_{\rm 
rest}$\,/\,\AA } & \colhead{redshift\,/\,$z$}} 
\startdata
6133.9   &  Ly-$\alpha$(em)\,$\lambda$\,1215.67 & 4.0457 \nl
6122.1   & Ly-$\alpha$(abs)\,$\lambda$\,1215.67 & 4.0360 \nl
6352.6 &    \ion{Si}{2}\,$\lambda$\,1260.4	& 4.0417 \nl
6551.9 &    \ion{O}{1}\,$\lambda$\,1302.2	& 4.0314 \nl
6563.4 &    \ion{Si}{2}\,$\lambda$\,1304.4	& 4.0317 \nl
6721.0\tablenotemark{1} &    \ion{C}{2}\,$\lambda$\,1334.5 & 4.0363 \nl
6730.6\tablenotemark{1} &    \ion{C}{2}\,$\lambda$\,1335.7 & 4.0390 \nl
7024.7 &    \ion{Si}{4}\,$\lambda$\,1393.8	& 4.0400 \nl
7069.2 &    \ion{Si}{4}\,$\lambda$\,1402.8	& 4.0393 \nl
7694.0 &    \ion{Si}{2}\,$\lambda$\,1526.7	& 4.0396 \nl
7805.6\tablenotemark{1} &   \ion{C}{4}\,$\lambda$\,1548.2	& 4.0417 \nl
7819.2\tablenotemark{1} &   \ion{C}{4}\,$\lambda$\,1550.8	& 4.0420 \nl
8109.6\tablenotemark{1} &   \ion{Fe}{2}\,$\lambda$\,1608.5	& 4.0417 \nl
8126.4 &   \ion{Fe}{2}\,$\lambda$\,1611.2	& 4.0437
\enddata
\tablenotetext{1}{Affected by a nearby sky emission features.}
\end{deluxetable}

\clearpage




\begin{deluxetable}{ccccccc}
\tablewidth{50pc} \tablecaption{Limits on high-ionization lines
\label{tab:noAGN}}
\tablehead{\colhead{Emission line} & \colhead{Rest-frame $\lambda$} & 
\colhead{Line flux\tablenotemark{1}} &
\colhead{Ly-$\alpha$ flux ratio\tablenotemark{1}} & 
\colhead{QSO ratio\tablenotemark{2}} &
\colhead{HzRG} & 
\colhead{Seyfert~II} \nl
\colhead{} & \colhead{\AA } & 
\colhead{$10^{-18}\,{\rm erg\,cm^{-2}\,s^{-1}}$} &
\colhead{$f$(Ly-$\alpha$)/$f$(line)} &
\colhead{$f$(Ly-$\alpha$)/$f$(line)} &
\colhead{ratio\tablenotemark{4}} &
\colhead{ratio\tablenotemark{5}} \nl
}
\startdata
Ly-$\alpha$ & 1215.7 & 37 & 1.0 & 1.0 & 1.0 & 1.0 \nl
\ion{N}{5} & 1238.8,1242.8 & $<$ 1.8 ($3\,\sigma$) & $>$ 20
($3\,\sigma$) & 4.0 & 9.7 & --- \nl
\ion{C}{4} & 1548.2,1550.8 & $<$ 5.0\tablenotemark{5} ($3\,\sigma$) & $>$
7.4\tablenotemark{5} ($3\,\sigma$) & 2.5 & 3.9 & 4.6 \nl 
\ion{He}{2} & 1640.5 & $<$ 1.4 ($3\,\sigma$) & $>$ 26 ($3\,\sigma$) & 20
& 5.5 & 27 \nl
\enddata
\tablenotetext{1}{These are $3\,\sigma$ upper limits in a
1\sq\arcsec\,$\times$\,7\,\AA\ aperture.}
\tablenotetext{2}{From the composite broad-line
region AGN spectrum in Osterbrock (1989).}
\tablenotetext{3}{From the composite high-redshift radio galaxy (HzRG)
spectrum of Stern \etal\ (1999) from the MG sample.}
\tablenotetext{4}{From Ferland \& Osterbrock (1986).}
\tablenotetext{5}{\ion{C}{4} lies in the OH
forest sky emission so the limit is less stringent.}
\end{deluxetable}

\clearpage

%

\begin{deluxetable}{ccccc}
\tablewidth{33pc} \tablecaption{Photometry of the arcs (integrated over
6\arcsec\ length for the northern arc and 3\arcsec\ for the
southern). Magnitudes are on the natural Vega system, and the
signal-to-noise ratio for each measurement is indicated.
\label{tab:arc_phot}}
\tablehead{\colhead{Component} & \colhead{F555W} &
\colhead{F814W} & \colhead{$H$} & \colhead{$K'$}}
\startdata
\nl
northern arc & $24.82$ ($\sigma=16$) & $22.61$ ($\sigma=57$) &
$21.19$ ($\sigma=20$) & $20.38$ ($\sigma=25$) \nl
southern arc & $25.14$ ($\sigma=17$) & $23.13$ ($\sigma=51$) &
$21.99$ ($\sigma=13$) & $21.16$ ($\sigma=17$) \nl
\enddata
\end{deluxetable}


\clearpage

\clearpage

\clearpage

\clearpage

\clearpage

\clearpage

\clearpage

\clearpage

\clearpage

\clearpage

\clearpage

\clearpage

\clearpage

\clearpage

\clearpage
\appendix
\section{Appendix A: The ``straight arc''}
\label{sec:straight_arc}

Our near-infrared imaging covers the nearby ``straight arc'' (Pell\'{o}
\etal\ 1991), which has generated much interest given the challenge to
lens models in reproducing this geometry (\eg, Kassiola, Kovner \&
Blandford 1992). However, our $H$ and $K'$ images indicate that the
``straight arc'' is in fact a serendipitous alignment of two or three
significantly less straight components (Fig.~\ref{fig15}). The
photometry of these components is detailed in
Table~5. Although the colors of A \& B are
similar, C differs significantly -- it is redder by a magnitude in
$(I_{814}-K')$ -- adding weight to the interpretation that these are
physically distinct objects. Indeed, spectroscopy by FB98 indicates that
the two components have different redshifts ($z_{A}=1.033$ \&
$z_{C}=0.913$).


\placefigure{fig15}

\setcounter{figure}{0}

\clearpage
\begin{figure}[ht]
\plotone{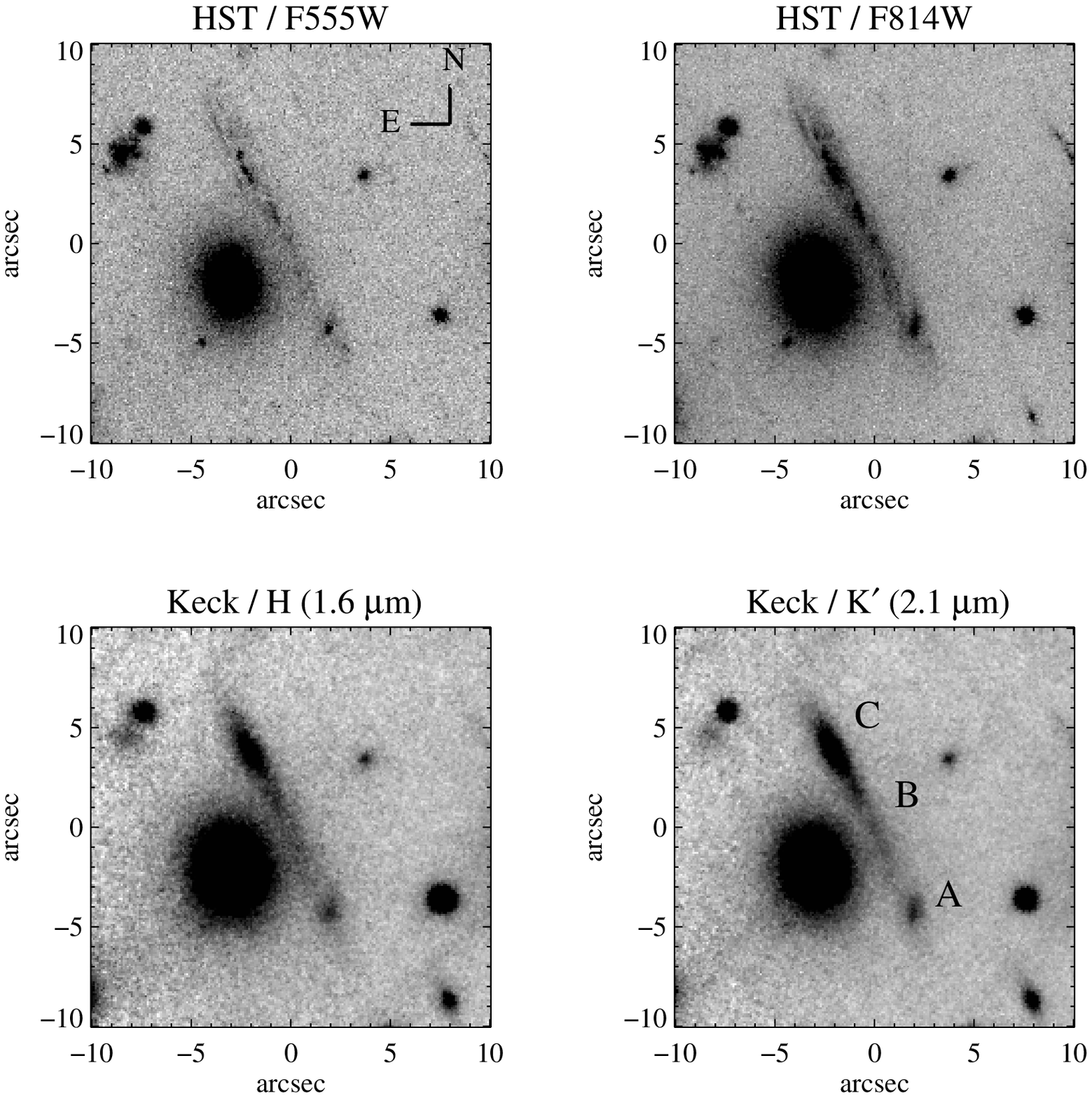}
\figcaption[fig15.eps]{The HST/optical and Keck/near-infrared imaging of the
``straight arc'' behind Abell~2390. The $H$ and $K'$ imaging clearly
shows that this is composed of two or three physically distinct objects,
one of which (C) exhibits much redder colors than A or B. \label{fig15}}
\end{figure}

\clearpage

%

\setcounter{table}{0}

\begin{deluxetable}{ccccc}
\tablewidth{33pc} \tablecaption{The colors on the Vega system of the
components of the ``straight arc'' lensed by Abell~2390. The
signal-to-noise ratio is typically $>100$ for each component in each
band, so the random errors are $\sim 0.01^{m}$, and the systematic
uncertainty in the photometric zeropoints dominates.
\label{tab:straight_arc_cols}}
\tablehead{\colhead{Component} & \colhead{$K'$} &
\colhead{$(V_{555}-K')$} & \colhead{$(I_{814}-K')$} & \colhead{$(H-K')$}}
\startdata
\nl
A & 19.23 & 1.67 & 0.67 & 0.59 \nl
B & 19.06 & 1.79 & 0.78 & 0.55 \nl
C & 18.00 & 2.77 & 1.72 & 0.92 \nl
\enddata
\end{deluxetable}


\end{document}